\documentclass[12pt]{article}
\usepackage{amsmath,amsfonts,amssymb}
 \usepackage{epsf,graphics,graphicx}

\newcommand{\be}{\begin{equation}}

\newcommand{\ee}{\end{equation}}
\newcommand{\bea}{\begin{eqnarray}}
\newcommand{\eea}{\end{eqnarray}}
\newcommand{\bee}{\begin{equation*}}
\newcommand{\eee}{\end{equation*}}

\newcommand{\myfig}[4]{
\begin{figure}[h]
\begin{center}
\includegraphics[scale=#2]{#1}
\end{center}
\caption{#3}
\label{fig:#4}
\end{figure}
}

\begin{document}

\title{Option Pricing, Historical Volatility and Tail Risks}
\author{Samuel E. V\'azquez \\
\small Baruch College, CUNY \\
samuel.vazquez@baruch.cuny.edu
 }

\maketitle

\abstract{
We revisit the problem of pricing options with historical volatility estimators. We do this in the context of a generalized GARCH model with multiple time scales and asymmetry. 
It is argued that the reason for the observed volatility risk premium is tail risk aversion. 
We parametrize such risk aversion in terms of three coefficients: convexity, skew and kurtosis risk premium. 
We propose that option prices under the real-world measure are not martingales, but that their drift is governed by such tail risk premia.
We then derive a fair-pricing equation for options and show that the solutions can be written in terms of a stochastic volatility model in continuous time and under a martingale probability measure. This gives a precise connection between the pricing and real-world probability measures, which cannot be obtained using Girsanov Theorem. We find that the convexity risk premium, not only shifts the overall implied volatility level, but also changes its term structure. Moreover, the skew risk premium makes the skewness of the volatility smile steeper than a pure historical estimate. We derive analytical formulas for certain implied moments using the Bergomi-Guyon expansion. This allows for very fast calibrations of the models. We show examples of a particular model which  can reproduce the observed SPX volatility surface using very few parameters.

}

\section{Introduction}
Most option pricing models are written directly in the martingale or pricing probability measure \cite{musiela}. Such models usually have a significant number of parameters which need to be fitted to the volatility surface. In the end, such parameters will show strong time dependency, invalidating the initial assumptions of the model. Moreover, the final values of the parameters have little physical significance, and so there is no notion of a ``fair" option price. We think of this as a fit-only approach, which in our opinion is best done in the context of parametric smile models such as SSVI \cite{ssvi}.

On the other hand, there is another stream of literature which studied the volatility surface produced by GARCH volatility forecasts \cite{duan,jacobs,engle,jacobs1}. However, here one runs into another problem: what is the relation between the real-world and the pricing or martingale probability measure? One solution is to leave free some of the GARCH parameters so they can be fitted to the volatility surface. However, this puts us back into the fit-only approach without any understanding of the physical meaning of these parameters. Worst, the GARCH models are written in discrete time, and hence require time-consuming Monte Carlo simulations in order to find the optimal parameters.

An early attempt to find a direct relation between the pricing and real-world measure in the context of GARCH models is found in \cite{duan}. This approach assumes that the one-period expected variance, is the same in both probability measures. However, this assumption is wrong as we will show in this paper. In fact, due to tail risks, there should be a significant premium paid to the short gamma trader even for  one-day returns. In other approaches, such as \cite{jacobs}, the authors start by modeling log returns in discrete time and define the pricing measure by requiring  simple returns to be a martingale.
However, as we will show in this paper, the volatility risk premium has nothing to do with the drift of the underlying. In fact, option prices are very insensitive to drifts and the underlying can very well be a martingale in both the real-world and pricing measure.  A notable exception to this literature stream is \cite{jacobs2,jacobs3}, which uses a price kernel approach to connect both probability measures. We believe there is an interesting connection between their approach and ours, but we will leave this for future work.

In this article we introduce a new approach to option pricing. Our goal is to use historical volatility estimators, while introducing risk premia which will allow to fit the volatility surface. 
In fact, only the risk premia needs to be fitted to the option prices. The rest of the parameters will be determined using the time series of the underlying. We will show that most of the features of the volatility surface can be explained using a good volatility forecast and only three risk premia. We will assume that the risk premia are constant. This is not true in practice, but a generalization is possible and will be left for future work. We will argue that the risk premium parameters we introduce are related to tail risk aversion, and come from the fact that option traders mark to market their books in discrete time and have limited capital. 

We will begin by working in discrete time, but assume the time step to be small enough so that we can expand option prices to second order in variations of the stochastic variables. This is what most option traders do in practice. Moreover, as most practitioners know, only the first few Greeks can be traded in the market due to liquidity constraints.  This approximation will allow us to make a connection with the more familiar continuous time stochastic volatility models.


We define tail risk as a typical large move of the underlying, not necessarily a catastrophic ``Black-Swan" event \cite{taleb}. However, we do not assign probabilities to such events. In practice, all market participants have limited capital, and must limit their leverage so that they can withstand such tail events. In fact, most brokers determine margin requirement precisely this way, using stress testing. What does this mean in practice? Suppose you are a trader who is short gamma. Under a large move of the underlying price $S$, you face a potentially large loss of size: $\lim_{|\delta S| \rightarrow \infty} \delta P = -\Gamma \delta S^2$, where $\Gamma > 0$ is the net gamma exposure. This is an unhedgable tail risk! In other words, there is a large tail risk asymmetry between a long-volatility and a short-volatility position. The short volatility trader must then put aside more capital than the long volatility counter-party. This is a cost of carry and so it is only fair that the short-volatility trader gets compensated by having a non-zero drift in his/her portfolio: $\mathbb{E}[\delta P] >0$. This very simple argument is the basis of our option pricing approach. In a nutshell, we propose that in the real-world measure, the drift of option returns is governed by the prices of tail risk.

We will restrict ourselves to a class of GARCH models with asymmetry and multiple time scales. However, our methodology can be applied to more general models, even those that include high-frequency volatility estimators \cite{hfvol}. We derive a generalized Black-Scholes equation under the real-world measure. Using Feynman-Kac theorem, we map the solutions to this equation to a stochastic volatility model in continuous time and under a martingale probability measure. This gives a precise mapping from the real-world to the pricing measure. However, this connection cannot be obtained using the standard Girsanov transformation.

Using the results of Bergomi and Guyon \cite{bergomi}, we derive approximate formulas for certain implied moments up to second order in the volatility of volatility (vol-of-vol). These moments can  be compared to the corresponding strip of options for fast calibration. Each risk premium is calibrated independently. 
In particular, we show that we can get the convexity/gamma risk premium by fitting the variance swap term structure. Moreover, the skew and kurtosis risk premia are obtained by fitting similar strip of options. Once the risk premia are calibrated, one can generate full volatility surfaces using Monte Carlo simulations. We show that the volatility surfaces obtained this way are close to what we observe in the market. 

We should stress that the goal of this paper is not to provide a comparative study of GARCH models or best estimation techniques. Our purpose is simply to introduce a new pricing methodology and give some examples. Therefore, we will not attempt to compare the fit quality of different models.

In section \ref{sec:tail} we will make the tail risk argument more precise and define the risk premia. In section \ref{sec:garch11} we study in detail the GARCH(1,1) model which serves to illustrate the main ideas. In section \ref{sec:mgarch} we generalize the GARCH model to include asymmetry and multiple time scales.
In section \ref{sec:bergomi} we derive approximate formulas for certain implied moments of the underlying returns. In section \ref{sec:calibration} we explain how the calibration is done using SPX option data. Moreover, we give examples of the volatility surfaces obtained from a particular GARCH model. We conclude in section \ref{sec:conclusion}.

\subsection{Notation}
We denote the price of the underlying asset by $S_t$, where $t$ is time measured in years. As usual, we assume that $S_t$ is the forward price, so that we can ignore dividends and interest rates. When working in discrete time we take a one day time step: $\delta t = 1/252$ (in years). Simple returns will be denoted by
\bee
\delta S_t := S_t-S_{t-\delta t}
\eee
In general, time subscripts denote stochastic time dependence while parenthesis denote smooth time dependence. For example, $x_t(T)$ is a smooth function of $T$ for fixed $t$. Moreover, all stochastic processes of the form $x_t$ are $t$-measurable in the sense that they depend on information up to time $t$.

The underlying return will be decomposed as follows:
\bee
r_t := \frac{\delta S_t}{S_{t-\delta t}}  = \sqrt{\delta t\, \nu_{t-\delta t}}\epsilon_t 
\eee
where $\epsilon_t$ is a {\it i.i.d.} noise with zero mean and unit standard deviation, and $\nu_t$ is the realized annualized variance. Note that we take the underlying to be a martingale under the real-world measure. However, adding a drift or taking log-returns instead has a negligible effect on the parameters of the model. We also find little evidence for skewness in the distribution of $\epsilon_t$. Therefore, we will assume that the distribution of $\epsilon_t$ is symmetric.

We will make ample use of exponential moving averages or EMAs. Our definition is the following:
\be
\label{ema}
\text{EMA}_L[x_t] = \left(1 - \frac{1}{L}\right)\text{EMA}_L[x_{t-\delta t}]  + \frac{1}{L} x_t
\ee
where $L$ is the time scale of the EMA in days, and $x_t$ is some random process.

The real-world probability measure is denoted by $\mathbb{P}$. The notation $\mathbb{E}_t[x_T]$ for $T \geq t$ means conditional expectation with information up to time $t$.
The pricing measure will be denoted by $\mathbb{P}^\star$ with similar notation for the conditional expectation: $\mathbb{E}^\star_t[x_T]$.

\section{Tail Risks}
\label{sec:tail}
The tail risk of an option trader follows from the non-linear dependency of options on the movements of the underlying asset. We consider tail scenarios parametrized by the normalized return $\epsilon_t$. For example, $\epsilon_t = \pm 3$ is a ``3-sigma" scenario. Moreover, we use the notation $\lim_{|\epsilon_t|\rightarrow \infty}$ to denote a large underlying move (not literally infinite). Basically, we think about typical scenarios of 3-5 sigma. These are not Black Swan events, as they happen quite often. However, they are large enough to cause substantial losses to option traders and trigger margin calls.

Suppose we have a portfolio $P^{(2)}$ with some gamma exposure such that, under a tail event we have:
\be
\label{P2}
 \lim_{|\epsilon_t| \rightarrow \infty} \delta P^{(2)}_t  = \epsilon_t^2 
\ee
where the superscript in $P^{(2)}_t$ indicates the asymptotic quadratic dependency on $\epsilon_t$.
As we discussed in the introduction, a trader with a short position in $P^{(2)}$ will be asked by the broker to put more margin than the one with a long position.
This is a cost of carry, because he/she could be investing this money somewhere else.
In order to compensate this trader, the profits and losses (P\&L) of $P^{(2)}$ must have a drift in the real-world measure:
\be
\label{P2drift}
\mathbb{E}_t\left[\delta P^{(2)}_{t+\delta t}\right] = -\lambda_2 
\ee
where we expect $\lambda_2 > 0$ on average. We call $\lambda_2$ the {\it convexity or gamma risk premium}. Note that we do not have to know any details about this portfolio, but only its asymptotic exposure to $\epsilon$.  In fact, the key assumption of this paper is that the form of such portfolio does not matter, and that any other portfolio, say $\tilde P^{(2)}$, with the same tail risk will have the same drift. In other words, {\it derivative markets only price tail risks and not ``daily" variance}. 

A simple example of a portfolio with gamma exposure is the front VIX future contract. In figure \ref{fig:vixpnl} we compare the cumulative P\&L of the front short VIX contract with those of the front long SPMINI. Both P\&Ls have been risk managed so that they have the same daily risk in a scale of 20 days\footnote{More precisely, let $R_t = F_t- F_{t-\delta t}$ be the daily P\&L of the future contract. The risk managed P\&L is given by $\tilde R_t = R_t/\sqrt{\text{EMA}_{20}[R_{t-\delta t}^2]}$.}. We can clearly see that the VIX future has a greater risk premium than the SPMINI for the same daily risk. However, it also has larger draw-downs. In figure \ref{fig:vixcondpnl} we show the residual VIX future P\&L conditioned on the SPMINI future P\&L\footnote{The residual P\&L is defined by $r_\text{VIX} -\beta r_\text{SPMINI}$, where $\beta := \text{Cov}[r_\text{VIX},r_\text{SPMINI}]/\text{Var}[r_\text{SPMINI}]$, where $r_\text{VIX},r_\text{SPMINI}$ are the risk-managed PnLs of the VIX and SPMINI contracts respectively. }. It is clear that the short VIX future has a gamma component that causes quadratically large losses for large movements of the SPMINI. This is the reason for the extra premium!

\myfig{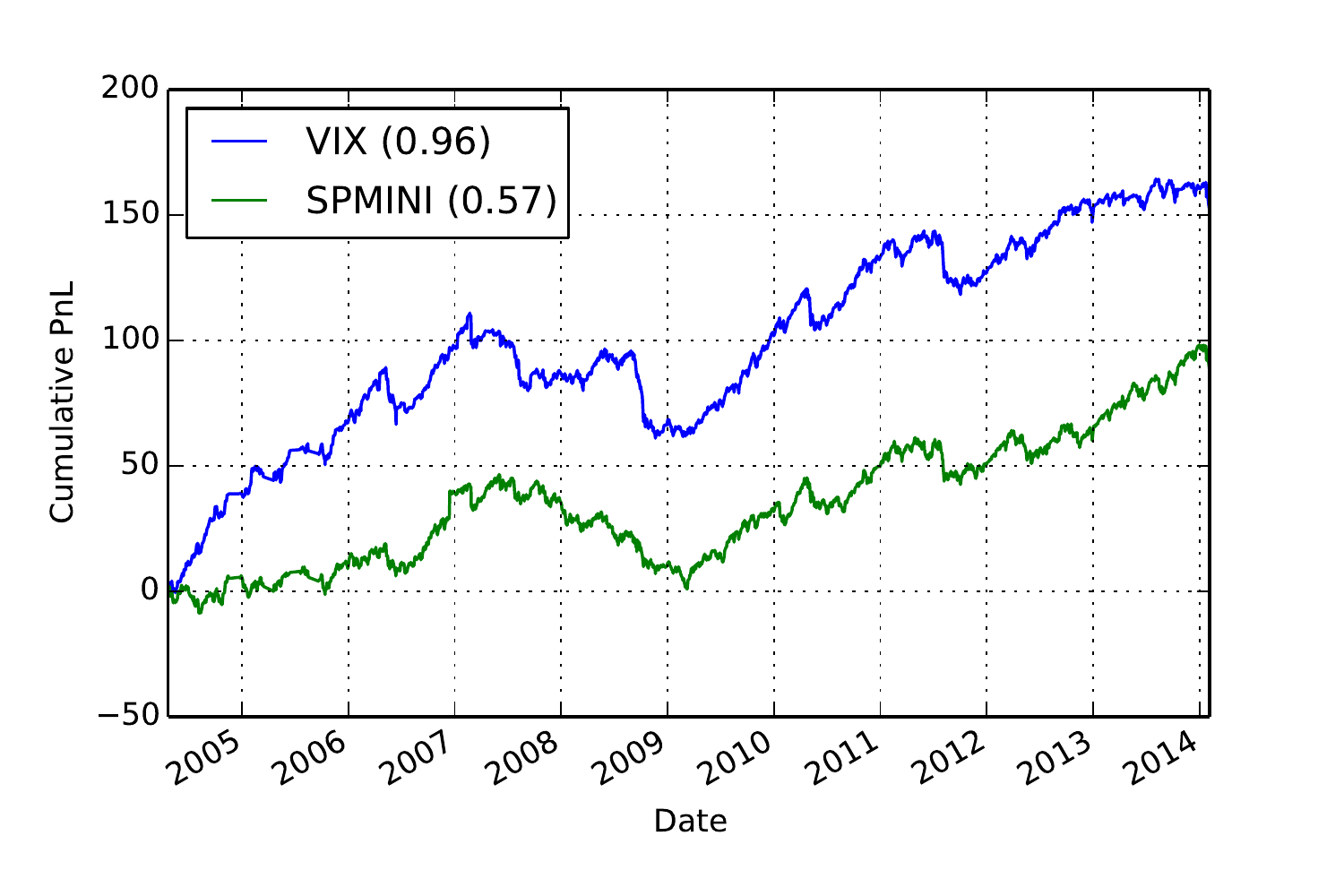}{0.6}{\small Cumulative P\&L of the front short VIX and long SPMINI futures. Each future has been risk-managed to maintain approximately one dollar of daily risk on a rolling scale of 20 days. The annualized Sharpe ratios are shown in parenthesis.  }{vixpnl}
\myfig{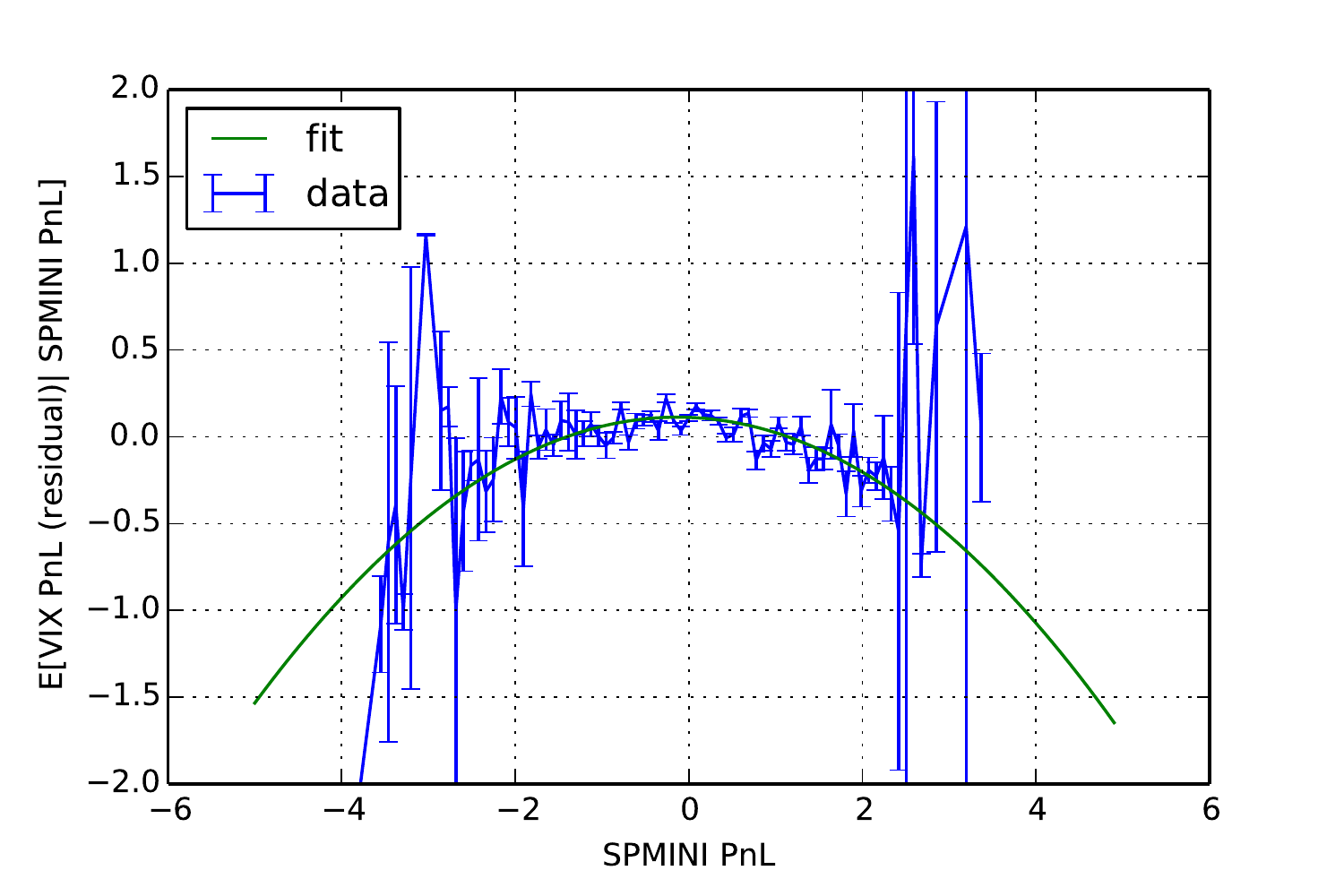}{0.6}{\small Front short VIX future residual P\&L conditioned on the front long SPMINI future P\&L.   The conditioning has been done by dividing the observations into 200 bins. We also show a quadratic polynomial fit for visual clarity.}{vixcondpnl}

Now consider a portfolio $P^{(3)}$ such that,
\be
\label{P3}
 \lim_{|\epsilon_t| \rightarrow \infty} \delta P^{(3)}_t  = \epsilon_t^3
\ee
In equity markets, most traders are afraid of the left tail. This means that the trader with a long position in $P^{(3)}$ is exposed to cubic losses under a large drawdown.  In such markets one expects to see a {\it skew risk premium} such that
\be
\label{P3drift}
\mathbb{E}_t\left[\delta P^{(3)}_{t+\delta t}\right] = \lambda_3
\ee
where $\lambda_3 > 0$ on average. In FX or certain commodity markets, we do not expect to see such risk premium as market participants are equally afraid to both the left and right tail. 

Note that this is a statement about risk aversion and not about the probability distribution of the market. In fact, one can argue that nobody know the true real-world  probability measure. However, all of us have capital requirements that become more stringent on downside equity markets (e.g. most investors are long equities by definition). 

Finally, we introduce a {\it kurtosis} risk premium:
\bea
\label{P4}
 \lim_{|\epsilon_t| \rightarrow \infty} \delta P^{(4)}_t  &=& \epsilon_t^4 \\
\label{P4drift}
\mathbb{E}_t\left[\delta P^{(4)}_{t+\delta t}\right] &=& -\lambda_4 
\eea
where we expect  $\lambda_4 > 0$ on average.

One can imagine higher moments, but as most option traders know, it is increasingly difficult to get such exposures due to liquidity constraints. The higher the moment, the more we need to leverage the option book and the less capacity there is for such strategy. Moreover, in the GARCH models studied below, we do not get higher order exposures if we restrict ourselves to second-order Greeks. 
The risk premia $(\lambda_2,\lambda_3,\lambda_4)$ will turn out to be the only parameters than need to be fitted to option prices.

\section{The GARCH(1,1) Model}
\label{sec:garch11}
In this section we  study in detail the GARCH(1,1) model. This is the simplest model of the GARCH family and will serve to illustrate the main ideas.
The goal of this section is to derive the pricing or martingale probability measure for this model using a tail risk argument. We begin by pricing a variance swap, and later move to price a general European contingency claim.

The GARCH(1,1) model is basically an EMA filter:
\bea 
\label{vswnu}
\nu_t &=& \bar \nu(1-\alpha) + \alpha X_t  \\
\label{vswx}
X_t &=& \frac{1}{\delta t} \text{EMA}_L[r_t^2] \\
\label{vswdx}
\delta X_{t+\delta t} &=& \frac{1}{L}\left(\nu_t \epsilon_{t+\delta t}^2 - X_t\right) 
\eea
where $\bar \nu$ is the unconditional variance and $\alpha \in [0,1]$ is a parameter that controls the strength of the volatility autocorrelation.

\subsection{Pricing a Variance Swap}
Let's now begin by pricing a variance swap contract with maturity date $T$. We denote the price of such contract at time $t$ by $V_t(T)$. At  expiry our variance swap pays
$P_T(T) = \sum_{j=1}^N r_{t+j\delta t}^2 - V_t(T)$, where $T = t+N\delta t$ is the expiry date and $N$ the number of days between $t$ and $T$. Since it takes zero capital to enter such contract, the P\&L of the variance swap between time $t$ and $t+\delta t$ is given by
\be
\label{vswpnl}
\delta P_{t+\delta t}(T) = V_{t+\delta t}(T)-V_t(T) + r_{t+\delta t}^2
\ee
We now assume that the price $V_t(T)$ is a smooth function of time and the filter $X_t$, $V_t(T) := V(t,X_t)$. Moreover, note that the boundary condition is $V(T,X) =0$.
Up to second order in variations of $X$ and assuming a small enough time step $\delta t$ we have,
\be 
\label{vapprox}
\delta P_{t+\delta t}(T) \approx \frac{\partial V}{\partial t} \delta t + \frac{\partial V}{\partial X_t}\delta X_{t+\delta t} +\frac{1}{2} \frac{\partial^2 V}{\partial X_t^2}\delta X_{t+\delta t}^2 + \nu_t \epsilon_{t+\delta t}^2 \delta t
\ee
This expansion will turn out to be exact in this case.

We now look at the tail risks of the variance swap. Using Eqs. (\ref{P2}), (\ref{P4}) and (\ref{vswdx}) in Eq. (\ref{vapprox}), we can decompose the asymptotic limit of the variance swap P\&L as follows:
\bea
\label{vswlim}
\lim_{|\epsilon_{t+\delta t}|\rightarrow\infty} \delta P_{t+\delta t}(T)  &=& \lim_{|\epsilon_{t+\delta t}|\rightarrow\infty} \left [\frac{\partial V}{\partial X_t}\frac{\nu_t}{L}\delta P_{t+\delta t}^{(2)}  + \nu_t\delta t \delta P_{t+\delta t}^{(2)} + \frac{1}{2} \frac{\partial^2 V}{\partial X_t^2} \frac{\nu_t^2 }{L^2} \delta P_{t+\delta t}^{(4)}\right]
\eea
We should emphasize that since we want to consider a general solution $V(t,X)$, we cannot compare the different terms in  Eq. (\ref{vswlim}) as we do not know the magnitude of the derivatives. In fact, for the variance swap it turns out that $V$ is a linear function of $X$ and so the second derivative vanishes.

According to our argument in the previous section, any two portfolios with the same tail risks should have the same drift.
Therefore, using Eqs. (\ref{P2drift}) and (\ref{P4drift}) and the asymptotics given in Eq. (\ref{vswlim}), we conclude that the drift of the variance swap must be given by
\bea 
\label{vswfair}
\mathbb{E}_t\left[\delta P_{t+\delta t}\right] &=& \mathbb{E}_t\left[\delta P_{t+\delta t}^{(2)} \right]\left(\frac{\partial V}{\partial X_t}\frac{\nu_t}{L} + \nu_t\delta t\right)+\frac{1}{2} \mathbb{E}_t\left[\delta P_{t+\delta t}^{(4)} \right] \frac{\partial^2 V}{\partial X_t^2} \frac{\nu_t^2 }{L^2} \nonumber \\
&=&-\lambda_2  \left(\frac{\partial V}{\partial X_t}\frac{\nu_t}{L} + \nu_t\delta t\right)- \frac{\lambda_4}{2} \frac{\partial^2 V}{\partial X_t^2} \frac{\nu_t^2 }{L^2}
\eea
This is the fair-value equation for the variance swap under the real-world probability measure. More explicitly, we can write Eq. (\ref{vswfair}) as a PDE for $V(t,X)$:
\be
\label{vswpde}
\frac{\partial V}{\partial t} + \theta\left[\nu(1+\lambda_2) - X\right]\frac{\partial V}{\partial X} + \frac{1}{2} \xi^2 \nu^2 \frac{\partial^2 V}{\partial X^2} + (1+\lambda_2)\nu = 0
\ee
where 
\bea
\label{nu}
\nu  &=& \bar \nu (1-\alpha) + \alpha X  \\
\label{theta}
\theta &=&  (\delta t L)^{-1} \\
\label{xi}
\xi &=& \frac{\sqrt{m_4-1 + \lambda_4}}{L \sqrt{\delta t}} \\
m_4 &=& \mathbb{E}[\epsilon^4]
\eea
 and the boundary condition is $V(T,X) = 0$. In writing Eq. (\ref{vswpde}) we have discarded a term quadratic in the drift of $\delta X$: $\mathbb{E}_t[\delta X_{t+\delta t}^2] \approx (m_4-1) \nu_t^2/L^2$. We find that, empirically, this is a very good approximation.

Using Feynman-Kac formula, one can write the solution to Eq. (\ref{vswpde}) in terms of a continuous time stochastic volatility process:
\bea
V_t(T) &=& (1 + \lambda_2) \int_t^T \mathbb{E}^\star_t[\nu_s] ds\nonumber \\
\nu_t  &=& \bar \nu (1-\alpha) + \alpha X_t \nonumber \\
dX_t &=& \theta[\nu_t(1+\lambda_2) - X_t]dt + \xi \nu_t dZ^\star_t \nonumber
\eea
Note that the pricing probability measure $\mathbb{P}^\star$ is just a mathematical trick to solve Eq. (\ref{vswpde}). However, it is very useful in order to get analytical solutions.
In fact, in this case the solution can be calculated explicitly:
\be
\label{vswprice}
V_t(T) = \bar X \tau + \alpha (1 +\lambda_2) \left(1 - e^{-\theta'\tau}\right)\frac{X_t - \bar X }{\theta'}
\ee
where the value of the filter $X_t$ is given by Eq. (\ref{vswx}) and
\bea
\tau    &=& T- t \nonumber \\
\theta' &=& \theta [1 -\alpha(1+\lambda_2)] \nonumber\\
\bar X  &=& \frac{\bar \nu (1-\alpha)(1+\lambda_2)}{1 - \alpha(1+\lambda_2)}\nonumber
\eea
Since Eq. (\ref{vswprice}) is linear in $X_t$, we can see that this is an exact solution to the variance swap price to all order in vol-of-vol. In fact, the solution only depends on the convexity or gamma risk premium. Therefore, by calibrating the variance swap term structure we can obtain the value of $\lambda_2$. Moreover, notice how the gamma risk premium not only shifts the level of the varswap, but also changes the effective mean-reversion time scale, which in turn changes the slope of the term structure.

Finally, note that the level of the implied expected variance is shifted from the historical one, even at the smallest time step:
\bee
\lim_{\delta t \rightarrow 0} \frac{V_t(t+\delta t)}{\delta t} = (1+\lambda_2)\nu_t
\eee
where $\nu_t$ is given by the historical estimate of Eq. (\ref{vswnu}).
This invalidates the assumption of \cite{duan}, who proposed that the one-period expected variance is the same in both the real-world and martingale probability  measures.

\subsection{Pricing Options}
We now generalize the previous problem to price a Europen-style option $C(t,S,X)$ with final payoff $C(T,S,X) = g(S)$. For a delta-hedged option, the second order expansion reads:
\be
\delta \hat C_{t+\delta t} \approx \frac{\partial C}{\partial t} \delta t + \frac{\partial C}{\partial X_t}\delta X_{t+\delta t} +\frac{1}{2} \frac{\partial^2 C}{\partial X_t^2}\delta X_{t+\delta t}^2 +  \frac{1}{2} \frac{\partial^2 C}{\partial S_t^2}\delta S_{t+\delta t}^2+\frac{\partial^2 C}{\partial S_t \partial X_t} \delta S_{t+\delta t} \delta X_{t+\delta t} 
\ee
where $\delta \hat C_{t+\delta t} = \delta  C_{t+\delta t} -\frac{\partial C}{\partial S_t} \delta S_{t+\delta t} - r C_t \delta t$ is the P\&L of the self-financed and delta-hedged option, and $r$ is the risk-free rate which we take to be constant.
The tail risks now include a skew contribution due to the cross term $\delta S\delta X \sim \epsilon^3$. More precisely, we have:
\bea
\lim_{|\epsilon_{t+\delta t}| \rightarrow \infty} \delta \hat C_{t+\delta t}  &=& \lim_{|\epsilon_{t+\delta t}| \rightarrow \infty}
\left[ \frac{\nu_t}{L} \frac{\partial C}{\partial X_t}\delta P_{t+\delta t}^{(2)} + \frac{\nu_t^2 }{2 L^2} \frac{\partial^2 C}{\partial X_t^2}  \delta P_{t+\delta t}^{(4)} + \frac{S_t^2 \nu_t \delta t}{2} \frac{\partial^2 C}{\partial S_t^2}   \delta P_{t+\delta t}^{(2)} \right. \nonumber \\
&& \left. + \frac{\sqrt{\delta t} \nu_t^{3/2} S_t}{L} \frac{\partial^2 C}{\partial S_t \partial X_t} \delta P_{t+\delta t}^{(3)}
\right]
\eea
Hence, the drift of the delta-hedged option is given by:
\bea
\mathbb{E}_t[\delta \hat C_{t+\delta t}]  &=&  -\lambda_2 \left( \frac{\nu_t}{L} \frac{\partial C}{\partial X_t} 
+ \frac{S_t^2 \nu_t \delta t}{2} \frac{\partial^2 C}{\partial S_t^2}
\right) + \lambda_3 \frac{\sqrt{\delta t} \nu_t^{3/2} S_t}{L} \frac{\partial^2 C}{\partial S_t \partial X_t} 
-\lambda_4 \frac{\nu_t^2 }{2 L^2} \frac{\partial^2 C}{\partial X_t^2}      
\eea
which leads to the following PDE for the option price:
\be
\label{garchpde}
\frac{\partial C}{\partial t} - r C  + \theta \left[ \nu(1+\lambda_2)-X\right] \frac{\partial C}{\partial X} +\frac{1}{2} (1+\lambda_2) \nu S^2 \frac{\partial^2 C}{\partial S^2} + \frac{1}{2} \xi^2 \nu^2 S^2 \frac{\partial^2 C}{\partial X^2} + \sqrt{1+\lambda_2} \rho \xi \nu^{3/2} S \frac{\partial^2 C}{\partial S\partial X} = 0
\ee
where $\nu$, $\theta$ and $\xi$ are defined in Eqs. (\ref{nu}) - (\ref{xi}) and
\be
\label{rho}
\rho = -\frac{\lambda_3}{\sqrt{(1+\lambda_2)(m_4-1+\lambda_4)}}  
\ee

Using Feynman-Kac formula, we can write the solutions to Eq. (\ref{garchpde}) in terms of the following stochastic process
\bea
\label{garchopt}
C_t(T) &=& e^{-r(T-t)} \mathbb{E}^\star_t\left[g(S_T)\right] \\
\label{garchds}
\frac{dS_t}{S_t} &=& \sqrt{(1+\lambda_2)\nu_t}  dW_t^\star \\
\label{garchnu}
\nu_t &=& \bar \nu (1-\alpha) + \alpha X_t \\
\label{garchx}
dX_t &=& \theta(\nu_t(1+\lambda_2) - X_t)dt + \xi \nu_t dZ_t^\star \\
\label{garchrho}
\mathbb{E}^\star[dW^\star_t dZ^\star_t] &=& \rho dt
\eea

As in the special case of the variance swap, the probability measure $\mathbb{P}^\star$ is the so-called martingale or pricing measure. 
It is interesting to note that we have derived a direct connection between the real-world and pricing measures parametrized by the three risk premia $(\lambda_2,\lambda_3,\lambda_4)$. These are the only parameters that must be inferred from the option prices. The rest is completely determined by historical data, including the initial value of the EMA filter $X_t$. 

Looking at Eq. (\ref{garchds}) we notice how the convexity risk premium $\lambda_2$ makes the implied volatility higher than the historical one (on average).
Moreover, we can see that this risk premium cannot be absorbed into the probability measure using a Girsanov transformation on $(W^\star,Z^\star)$. There is a fundamental reason for this: the presence of $\lambda_2$ comes from the fact that the option P\&L is marked to market in {\it discrete time}. Another way to see this is that the underlying price is a martingale both in the real-world and pricing measures. Therefore, the volatility risk premium has nothing to do with the drift of the underlying. Many authors seem to confuse the volatility risk premium with the equity risk premium. Those are two completely different quantities. In fact, there are many assets which do not have any obvious risk premium (e.g. FX rates or some commodities). However, their options still show a volatility risk premium. Therefore, any attempt to derive the pricing measure by putting a martingale condition on $S_t$ is doomed to fail (see e.g. \cite{duan,jacobs1}).

The skew risk premium $\lambda_3$ makes the correlation between the spot and the volatility more negative. In fact, even if the underlying distribution is symmetric, we can still have non trivial {\it implied} leverage effect due to the skew risk premium.  Finally, the kurtosis risk premium makes the implied vol-of-vol higher than the historical estimate.

The stochastic process given by Eqs. (\ref{garchds})  - (\ref{garchrho}) is well defined only if the risk premia obey the following bounds:
\bea
\label{garchbound1}
\lambda_2 &>& -1 \\
\label{garchbound2}
\lambda_4 &\geq& -m_4+1 + \frac{\lambda_3^2}{1+\lambda_2}
\eea
The second bound comes from the fact that we need $|\rho| \leq 1$. 

\section{Including Asymmetry and Multiple Time Scales}
\label{sec:mgarch}
There is a considerable number of studies that give evidence of multiple time scales in volatility auto-correlations (see for example \cite{volcorr1,volcorr2,volcorr3,volcorr4,volcorr5,volcorr6,volcorr7}) . In fact, it has been argued that volatility auto-correlations decay as a power law \cite{volcorr7}. One problem with a power-law filter is that it is non-Makovian. However, as shown in  \cite{filters}, one can always approximate a power law filter with multiple exponentials. Hence, in this section we study a generalized GARCH models which is a linear combination of EMA filters with different time scales. 

Another stylized fact of volatility, is the so-called leverage effect \cite{leverage1}. In other words, for equity indices, negative returns tend to increase future volatility more than positive ones. In the context of GARCH models, this is captured by adding a filter that depends only on past negative returns \cite{leverage2}. Hence, we will study the following general class of models:
\bea 
\label{dmgarchr}
r_t &=& \sqrt{\nu_{t-\delta t}} \epsilon_t \sqrt{\delta t} \\
\label{dmgarchnu}
\nu_t &=& \sum_{i=1}^{N+M} \alpha_i X^i_t   \\
\label{dmgarchx}
 X^i_t &=& \begin{cases}
\frac{1}{\delta t} \text{EMA}_{L_i}[r_t^2] &  \text{for}\; i = 1,\ldots,N \\
\frac{2}{\delta t} \text{EMA}_{L_i}[r_t^2 \mathbf{1}_{r_t < 0}] & \text{for}\; i = N+1,\ldots,N+M
\end{cases}
 \eea
 where $\sum_{i=1}^{N+M} \alpha_i = 1$, $\delta t = 1/252$, $r_t = S_t/S_{t-\delta t}-1$ , and the {\it i.i.d.} noise term $\epsilon_t$ has zero mean and unit standard deviation. 
Note that we do not have constant unconditional variance in Eq. (\ref{dmgarchnu}) as we did in the simple GARCH(1,1) model. However, we can always take one of the time scales to infinity, say $L_1 \rightarrow \infty$. This way we can recover the usual GARCH(1,1) model for example.
In practice we will take $L_1 = 1000$ days. This way we avoid too much in-sample bias as we only use past observations and we avoid having to fit the long term unconditional variance.

In order to find the pricing measure for this model, we can go over the same argument as in section \ref{sec:mgarch}. However, when expanding the option P\&L we now will have the following new tail risks:
\bea
\lim_{|\epsilon_t| \rightarrow\infty} \delta X_{t} \propto \epsilon_{t}^2 \mathbf{1}_{\epsilon_{t} < 0} \\
\lim_{|\epsilon_t| \rightarrow\infty}\delta S_{t} \delta X_{t} \propto \epsilon_{t}^3 \mathbf{1}_{\epsilon_{t} < 0} \\
\lim_{|\epsilon_t| \rightarrow\infty} \delta X_{t}^2 \propto \epsilon_{t}^4 \mathbf{1}_{\epsilon_{t} < 0} 
\eea
where $\delta X_t$ is one of the asymmetric filters.
We can now imagine ideal portfolios,  so that
\bee
 \lim_{|\epsilon_t| \rightarrow \infty} \delta \tilde P^{(n)}_{t}  = \epsilon_{t}^n \mathbf{1}_{\epsilon_{t} < 0} 
\eee
for $n=2,3,4$. In order to avoid introducing more risk premia for our model, we will argue that in equity markets, investors are only afraid of large {\it negative} returns. In other words, they only value downside tail risk. Therefore, these new tail risks must have the same drift as the symmetric ones:
\bea
\mathbb{E}_t[ \delta \tilde P^{(2)}_{t+\delta t}]  &=& \mathbb{E}_t[ \delta  P^{(2)}_{t+\delta t}] = -\lambda_2 \\
\mathbb{E}_t[ \delta \tilde P^{(3)}_{t+\delta t}]  &=& \mathbb{E}_t[ \delta  P^{(3)}_{t+\delta t}] = \lambda_3 \\
\mathbb{E}_t[ \delta \tilde P^{(4)}_{t+\delta t}]  &=& \mathbb{E}_t[ \delta  P^{(4)}_{t+\delta t}] = -\lambda_4 
\eea
where we used Eqs. (\ref{P2drift}), (\ref{P3drift}) and (\ref{P4drift}).

In order to value an option, we assume as before that its price is a smooth function of time, the spot and the filters: $C_t(T) = C(t,S,X)$. Expanding to second order in variations and taking into account the tail risks as in the previous section, we get the following PDE:
\bea
\label{mgarchpde}
&&\frac{\partial C}{\partial t} - r C  + \sum_i \theta_i \left[ \nu \delta_i -X^i\right] \frac{\partial C}{\partial X^i} +\frac{1}{2} (1+\lambda_2) \nu S^2 \frac{\partial^2 C}{\partial S^2} + \frac{1}{2} \sum_{ij} \xi_i \xi_j \rho_{ij}  \nu^2 S^2 \frac{\partial^2 C}{\partial X^i \partial X^j} \nonumber \\
&&+ \sqrt{1+\lambda_2} \sum_i \rho_i \xi_i \nu^{3/2} S \frac{\partial^2 C}{\partial S\partial X^i} = 0
\eea
where we have dropped terms quadratic in the drift of $\delta X_t^i$ and we have defined the following variables:
\bea 
\nu &=& \sum_i \alpha_i X^i \\
\theta_i &=& (L_i\delta t)^{-1} \\
\delta_i &=& 
\begin{cases}
1+\lambda_2 &  \text{for}\;\; i = 1,\ldots,N \\
1+2\lambda_2 & \text{for} \;\;i = N+1,\ldots,N+M
\end{cases} \\
\xi_i &=& 
\begin{cases}
\frac{\sqrt{m_4-1+ \lambda_4}}{L_i \sqrt{\delta t}} & \text{for}\;\; i = 1,\ldots,N \\
\frac{\sqrt{2m_4-1 + 4 \lambda_4}}{L_i \sqrt{\delta t}} & \text{for} \;\;i = N+1,\ldots,N+M
\end{cases} \\
\rho_i &=& 
\begin{cases}
-\frac{\lambda_3}{\sqrt{(1+\lambda_2)(m_4-1 + \lambda_4)}} & \text{for}\;\; i = 1,\ldots,N \\
\frac{2(m_3^- -  \lambda_3)}{\sqrt{(1+\lambda_2)(2m_4 -1+ 4\lambda_4)}} & \text{for} \;\;i = N+1,\ldots,N+M
\end{cases} \\
m_3^- &=& \mathbb{E}[\epsilon^3\mathbf{1}_{\epsilon <0}] \\
m_4 &=& \mathbb{E}[\epsilon^4] 
\eea
Moreover, the correlation between the filters is one if both are symmetric or asymmetric ($\rho_{ij} =1$), but the correlation between 
a symmetric and asymmetric filter is:
\bea
\rho_{ij} = \frac{m_4-1 + 2\lambda_4}{\sqrt{(m_4-1+\lambda_4)(2m_4-1+4\lambda_4)}} \;,
\eea
for $ i \in \{1,\ldots,N\}\;, j \in \{N+1,\ldots,N+M\}$. 

Using Feynan-Kac formula, we can relate the solutions of Eq. (\ref{mgarchpde}) to the following stochastic volatility model:
\bea
\label{mgarchopt}
C_t(T) &=& e^{-r (T-t)} \mathbb{E}^\star_t\left[g(S_T)\right] \\
\label{mgarchds}
\frac{dS_t}{S_t} &=&  \sqrt{(1+\lambda_2) \nu_t} dW_t  \\
\label{mgarchnu}
\nu_t &=& \sum_i \alpha_i X^i_t  \\
\label{mgarchdx}
dX^i_t &=& \theta_i[\nu_t\delta_i - X^i_t]dt + \xi_i \nu_t dZ^i_t \\
\label{mgarchrhoi}
\mathbb{E}^\star[ dW_t dZ_t^i] &=& \rho_i dt  \\
\label{mgarchrhoij}
\mathbb{E}^\star[dZ^i_t dZ_t^j] &=& \rho_{ij} dt 
\eea
The Brownian motions can be decomposed into a few PCA factors as follows:
\bea
\label{dZ}
dZ^i_t &=& 
\begin{cases}
\rho_+ dW_t + \sqrt{1-\rho_+^2}\left( \sqrt{|\bar \rho_{+-}|} d Z_t + \sqrt{1-|\bar\rho_{+-}|} d Z^+_t\right) & \text{for}\; i = 1,\ldots,N \\
\rho_- dW_t + \sqrt{1-\rho_-^2}\left( \text{sign}(\bar \rho_{+-})\sqrt{|\bar \rho_{+-}|} d Z_t + \sqrt{1-|\bar\rho_{+-}|} d Z^-_t\right) &  \text{for}\; i = 1+N,\ldots,N+M 
\end{cases}
\nonumber \\
\eea
where all Brownian motions on the RHS  $(W, Z, Z^+,Z^-)$ are  uncorrelated, and
\bea
\rho_+ &=&  -\frac{\lambda_3}{\sqrt{(1+\lambda_2)(m_4-1 + \lambda_4)}} \\
\rho_- &=& \frac{2(m_3^- -  \lambda_3)}{\sqrt{(1+\lambda_2)(2m_4 -1+ 4\lambda_4)}} \\
\rho_{+-} &=& \frac{m_4-1 + 2\lambda_4}{\sqrt{(m_4-1+\lambda_4)(2m_4-1+4\lambda_4)}} \\
\bar \rho_{+-} &=& \frac{\rho_{+-}-\rho_+\rho_-}{\sqrt{(1-\rho_+^2)(1-\rho_-^2)} }
\eea
Consistency of the model requires the following constraints:
\bea
\label{mgarchcond1}
|\rho_+| &\leq& 1 \\
\label{mgarchcond2}
|\rho_-| &\leq& 1 \\
\label{mgarchcond3}
|\rho_{+-}| &\leq& 1 \\
\label{mgarchcond4}
|\bar\rho_{+-}| &\leq& 1 
\eea
Note that condition (\ref{mgarchcond4}) is the most stringent bound. This condition translates into a minimum bound for the kurtosis risk premium:
\be
\label{bound}
\lambda_4 \geq \frac{4(m_4-1)(m_3^--\lambda_3) m_3^- + \lambda_3^2(2m_4-1)-m_4(m_4-1)(1+\lambda_2)}{(2m_4-1)(1+\lambda_2) - 4 (m_3^-)^2}
\ee
Note that Eq. (\ref{bound}) only applies for the general case where we have both symmetric and asymmetric filters. If all filters are symmetric ($M=0$) we only need to enforce Eq. (\ref{mgarchcond1}). On the other hand, if all filters are asymmetric ($N=0$), we only need to impose Eq. (\ref{mgarchcond2}).

Empirically, we find that Eq. (\ref{bound}) is saturated most of the time. The saturation of this condition can be interpreted as saying that, in this class of models, volatility has only one risk factor (apart from the spot moves). This makes sense, as in the discrete model, all filters are driven by the spot returns (squared). It would be interesting to generalize the model so that we generate more volatility risk factors. This can be done, for example, by adding a high-frequency filter to our volatility estimate.

To conclude this section, we will price forward variance and a variance swap for the model of Eqs. (\ref{mgarchds}) - (\ref{mgarchrhoij}). We will make ample use of these results in the next section.
We begin by introducing the matrix:
\bee \Omega_{ij} := \theta_i\left(\delta_{ij} - \delta_i\alpha_j \right) \eee
and its eigenvalue decomposition,
\bee \Omega = U\cdot D\cdot U^{-1}\;,\;\;\; D_{ij} := \tilde \theta_i \delta_{ij}
\eee
Forward variance is defined as 
\be
F_t(T) := \mathbb{E}_t^\star [\nu_T]
\ee
For our multi-scale model, we get
\bea
\label{fwd}
F_t(T) &=&  (1+\lambda_2) \sum_i \tilde \alpha_i \tilde X^i_t e^{-\tilde \theta_i (T-t)} \nonumber \\
\tilde \alpha &:=& U^T\cdot \alpha \nonumber \\
\tilde X_t &:=& U^{-1} \cdot X_t\nonumber
\eea
Moreover, note that forward variance is a martingale:
\be
\label{dfwd}                                                                       
dF_t(T) = \nu_t \sum_{ij} \tilde \alpha_i (U^{-1})_{ij} \xi_{ij} e^{-\tilde \theta_i (T-t)} dZ^j_t
\ee
The variance swap is imply the integrated forward variance:
\bea
V_t(T) &:=& \int_t^T ds F_t(s) \nonumber \\
&=& (1+\lambda_2) \sum_i \frac{\tilde \alpha_i \tilde X^i_t}{\tilde \theta_i} \left(1- e^{-\tilde \theta_i (T-t)}\right)
\eea

\section{The Bergomi-Guyon Expansion}
\label{sec:bergomi}
In order to find the value of the risk premia $\lambda_i$, we need to fit our stochastic volatility model to option prices. However, our model does not have an analytical solution, so we will employ the perturbative expansion developed by Bergomi and Guyon in \cite{bergomi}. This is basically a vol-of-vol expansion (e.g. an expansion in $\xi$). In this section we will derive closed-form formulas for various implied moments which will be used in the next section to calibrate our models. This will avoid the use of Monte Carlo simulations in the calibration. We will not address the accuracy of the expansion. 

The main result of \cite{bergomi}, is that to second order in vol-of-vol, option prices can be approximated by certain functionals of the initial forward variance curve. More precisely, let $x=\log S_0$ be the log-price of the underlying at the initial time $t=0$, and $C^{(0)}$ be option price evaluated at zero vol-of-vol ($\xi=0$). Moreover, we take $T$ to be the time to expiry. Then, to second order in vol-of-vol, we can approximate the option price as
\bea
\label{capprox}
C(x,T) &\approx& 
\left[
1 + \frac{1}{2} C^{x f} \partial_x^2 (\partial_x -1) + \frac{1}{8} C^{ff} \partial_x^2 (\partial_x -1)^2 \right. \nonumber \\
&&\left.+ \frac{1}{8} (C^{xf})^2 \partial_x^4 (\partial_x-1)^2 + \frac{1}{2} C^\mu \partial_x^3(\partial_x-1)
\right] C^{(0)}(x,T)
\eea
where
\bea 
C^{(0)}     &=& \exp\left[\frac{1}{2} V \partial_x(\partial_x-1) \right] g(x) \\
\label{cxf}
C^{x f} &=& \int_0^T dt \int_t^T du \frac{\mathbb{E}^\star_0[dx_t dF_t(u)]}{dt} \\
\label{cff}
C^{ff} &=& \int_0^T dt \int_t^T ds \int_t^T du \frac{\mathbb{E}^\star_0[dF_t(s) dF_t(u)]}{dt} \\
\label{cmu}
C^{\mu} &=& \int_0^T dt \int_t^T du \frac{\mathbb{E}^\star_0[dx_t dF_t(u)]}{dt}  \frac{\delta C^{x f}}{\delta F_0(u)} \\
V &=&  \int_0^T dt F_0(t)
\eea
and $g(x)$ is the option payoff at expiry. Note that all integrals are functionals of the initial forward variance curve $F_0(s)$, and their functional derivative is defined such that:
\bee
\frac{\delta F_0(s)}{\delta F_0(u)} = \delta(u-s)
\eee
where $\delta(x)$ is the Dirac delta function.

One useful special case of Eq. (\ref{capprox}) is the moment generating  function, which can be derived by using the following payoff: $g(x) = e^{\alpha x}$.
We have,
\bea
M_T(\alpha ) &:=& \mathbb{E}^\star_0[e^{\alpha (x_T-x)}] \\
 &=& e^{\psi(\alpha)}
\eea
where
\bea
\psi(\alpha) \approx \frac{1}{2} \alpha (\alpha-1) V + \frac{1}{2} C^{x f} \alpha^2(\alpha-1) + \frac{1}{8} C^{ff} \alpha^2 (\alpha -1)^2 + \frac{1}{2} C^\mu \alpha^3(\alpha-1)
\eea
Using Eqs. (\ref{mgarchds}) and  (\ref{dfwd}) it is straightforward to evaluate the integrals (\ref{cxf}) - (\ref{cmu}):
\bea
\label{cxffinal}
C^{xf}(T) &\approx& \sum_i A_i  I_i^{xf}(T)  \\
\label{cfffinal}
C^{ff}(T) &\approx& \sum_i B_{ij}  I_{ij}^{ff}(T)  \\
\label{cmufinal}
C^{\mu}(T) &\approx& \sum_i \tilde \theta_i A_i A_j    I_{ij}^{\mu}(T)  
\eea
where\footnote{In deriving Eqs. (\ref{cxffinal}) - (\ref{cmufinal}) we have used the following approximation: $(1+\lambda_2)^n \mathbb{E}^\star_0[ \nu_T^n] \approx [F_0(T)]^n$. Any corrections to this approximation will lead to at least cubic order corrections in $\xi$ to the equations above.}
\bea
\label{A}
A_i &=& \frac{\tilde \alpha_i}{\tilde \theta_i} \sum_j   (U^{-1})_{ij} \xi_j \rho_j  \\
\label{B}
B_{ij} &=& \frac{\tilde \alpha_i \tilde \alpha_j}{\tilde \theta_i \tilde \theta_j} \sum_{k,l}   (U^{-1})_{ik}  (U^{-1})_{il} \xi_k \xi_l \rho_{kl}  \\
I_i^{xf}(T) &=& \int_0^T dt F_0^{3/2}(t) \left(1 - e^{-\tilde\theta_i(T-t)}\right)  \\
I_{ij}^{ff}(T) &=& \int_0^T dt F_0^2(t) \left(1 - e^{-\tilde\theta_i(T-t)}\right) \left(1 - e^{-\tilde\theta_j(T-t)}\right)  \\
 I_{ij}^{\mu}(T) &=& \frac{3}{2} \int_0^T dt F_0^{3/2}(t) \int_t^T du F_0^{1/2}(u) e^{-\tilde\theta_i(u-t)}\left(1 - e^{-\tilde\theta_j(T-u)}\right) 
\eea

It is important to note that the dependency of the skew and kurtosis risk premia $(\lambda_3,\lambda_4)$ is fully contained in the vector $A_i$ and the matrix $B_{ij}$, and that the integrals $I^{xf},I^{ff},I^\mu$ contain all the term structure dependency.
This means that, once we calibrate the convexity risk premium $\lambda_2$ using varswaps, we only have to evaluate the integrals one time. Another important observation is that $C^{xf}$ and $C^\mu$ only depend on $\lambda_2$ and $\lambda_3$. This is because the product $\rho_i\xi_i$ is independent of the kurtosis risk premium. This property is very useful for calibration as we will see below.

For model calibration, it is very useful to calculate the following normalized moments
\bea
\label{moment1}
\sqrt{\frac{-2{\cal M}_1}{T}} &\approx& \sqrt{\frac{V}{T}}  \\
\label{moment2}
\frac{2 {\cal M}_3}{\sqrt{T} (-2{\cal M}_1)^{3/2}} &\approx& \frac{1}{\sqrt{T} V^{3/2}}\left( C^{xf} + C^\mu\right)   \\
\label{moment3}
\frac{2 {\cal M}_3 +{\cal M}_2 - {\cal M}_1^2 + 2 {\cal M}_1 }{\sqrt{T} (-2{\cal M}_1)^{5/2}} &\approx& \frac{1}{\sqrt{T} V^{5/2}}\left( C^\mu + \frac{1}{4} C^{ff}\right)   
\eea
where
\bea
{\cal M}_1 &:=& \mathbb{E}_0^\star\left[\log(S_T/S_0)\right] = M_T'(0)  \\
{\cal M}_2 &:=& \mathbb{E}_0^\star\left[\log^2(S_T/S_0)\right]  = M_T''(0)  \\
{\cal M}_3 &:=& \mathbb{E}_0^\star\left[\left(S_T/S_0+1\right)\log(S_T/S_0)\right] = M_T'(1) + M_T'(0) 
\eea
Using the well known results of Madam and Carr \cite{carr}, we replicate the moments ${\cal M}_i$ using OTM options as follows:
\bea
\label{optmoment1}
{\cal M}_1 &=&  - e^{r T} \left(  \int_0^{S_0} \frac{dK}{K^2} P(K) + \int_{S_0}^\infty \frac{dK}{K^2} C(K) \right)  \\
\label{optmoment2}
{\cal M}_2 &=& 2 e^{r T} \left[  \int_0^{S_0} \frac{dK}{K^2} \left(1-\log(K/S_0)\right) P(K) + \int_{S_0}^\infty \frac{dK}{K^2}\left(1-\log(K/S_0)\right) C(K) \right]  \\
\label{optmoment3}
{\cal M}_3 &=&  e^{r T} \left[  \int_0^{S_0} \frac{dK}{K^2} \left(\frac{K}{S_0}-1\right) P(K) + \int_{S_0}^\infty \frac{dK}{K^2}\left(\frac{K}{S_0}-1\right) C(K) \right] 
\eea
where the integrals go over the option strikes, and calls and puts are denoted by $C(K)$ and $P(K)$ respectively.
This means that the LHS of Eqs. (\ref{moment1}) - (\ref{moment3}) can be calculated using option prices, while the RHS is given by our model. This is how we will find the risk premia $(\lambda_2,\lambda_3,\lambda_4)$ in the next section. We also note that Eq. (\ref{moment1}) only depends on the convexity risk premium $\lambda_2$ and Eq. (\ref{moment2}) only depends on $\lambda_2$ and $\lambda_3$. Therefore, the calibration can be done sequentially: we first calibrate the variance swap term structure using Eq. (\ref{moment1}) to get $\lambda_2$. Next we find $\lambda_3$ using Eq. (\ref{moment2}). Finally, the kurtosis risk premium $\lambda_4$ can be found using Eq. (\ref{moment3}).

Another interesting observation concerns the ATM volatility skew, defined by
\bee {\cal S}(T) := \left. \frac{\partial \sigma_\text{BS}(K,T)}{\partial \log K}\right|_{K = S_0}
\eee
where $\sigma_\text{BS}(K,T)$ is the Black-Scholes implied volatility.
To first, order in vol-of-vol we can write (see \cite{bergomi}):
\be
{\cal S}(T)  \approx \frac{C^{xf}}{2 \sqrt{T} V^{3/2}}
\ee
As we saw in the previous section, the skew risk premium tends to make the leverage correlations $\rho_i$ more negative.
Therefore, looking at Eqs. (\ref{cxffinal}) and (\ref{A}) we can see that the skew risk premium will make $C^{xf}$ more negative and hence the ATM
skew steeper (more negative) than the historical estimate with $\lambda_3=0$.

\section{Calibration}
\label{sec:calibration}
In this section we explain our calibration methodology and show some examples of the resulting volatility surfaces obtained with a particular GARCH model. Our purpose is not to find which  model is best, so we will concentrate on a simple one which incorporates both multiple scales and asymmetry. Namely, we take Eqs. (\ref{dmgarchr}) - (\ref{dmgarchx}) with $N = 2$, $M=1$ and $L_1 \rightarrow \infty$. In other words, for calibration purposes, we take $X^1_t$ to be a constant\footnote{We find that, for calibration purposes, the maximum likelihood optimization converges much faster if we avoid including a long EMA filter. Later we will use a 1000 day EMA to substitute the unconditional variance.}.

Calibration proceeds in two steps. First, one must fit the discrete-time GARCH model using the daily time series of the underlying. This determines all parameters with the exception of the risk premia $(\lambda_2,\lambda_3,\lambda_4)$. The latter are obtained by fitting the moments given in Eqs. (\ref{moment1}) - (\ref{moment3}) using OTM option prices, while taking into account the bound given in Eq. (\ref{bound}). This is, of course, only an approximation. In fact, there are two approximations: first we have expanded to second order in vol-of-vol and second we will need to approximate the infinite integrals of Eqs. (\ref{optmoment1}) - (\ref{optmoment3}) with a discrete set of strikes. For a more accurate calibration one must do Monte Carlo simulations of the stochastic processes, but this can be very time consuming.
We find that using our approximate method gives reasonable smile fits. 

In order to calibrate the GARCH model, we take daily data from 28 global equity indices from 1990-01-01 to 2012-12-31.
We assume universality in the sense that the normalized returns $\tilde r^\alpha_t := r^\alpha_t/\text{Std}[r^\alpha_t]$ all follow the same GARCH model with the same parameters and $X^1_t = 1$. This way we avoid having to estimate the long-term mean variance, which is a very noisy quantity, and we do not expect it to be universal\footnote{Some equity indices can naturally have more volatility as they might be composed of fewer stocks or come from countries which are perceived to be riskier than the US.}. Later on, we will take $X^1_t$ to be a 1000 days EMA in order to avoid too much in-sample bias. We are thus left with 4 parameters: $(\alpha_2,\alpha_3,L_2,L_3)$.
The maximum likelihood fit is done assuming that the innovations $\epsilon$ follows a Gaussian distribution with $\mathbb{E}[\epsilon] = 0$ and $\mathbb{E}[\epsilon^2] = 1$.
The function to minimize is the average of the individual likelihood functions:
\bee
L = \sum_{\alpha = 1}^{\cal N} \frac{1}{n_\alpha} \sum_{t=1}^{n_\alpha} \left[\frac{1}{2}\log(\nu_{t-1}^\alpha) - \log\left(\rho\left(\tilde r^\alpha_t/\sqrt{\nu^\alpha_{t-1}}\right)\right)\right]
\eee
where $n_\alpha$ is the number of observations for the equity index $\alpha$, ${\cal N}$ the number of time series, and $\rho$ is the distribution function of $\epsilon$.
We find very little difference in the parameters if we use log returns or if we use some other distribution such as Student-t. We find the following parameters: $(\alpha_2,\alpha_3,L_2,L_3) \approx (0.4,0.5,36,6)$. It is interesting to note that the mean-reversion time scale for the asymmetric filter, $L_3$, is much smaller than for the symmetric one, $L_2$. This means that the model reacts faster to negative returns.

In order to calculate the risk premia, we need to evaluate the integrals over OTM options in Eqs. (\ref{optmoment1}) - (\ref{optmoment3}).  This is done by taking SPX options\footnote{Option data is provided by OptionMetrics.} with $|\Delta| \in [0.01,0.5]$.
The integrals are then approximated using the trapezoidal rule:
\bea
\int_a^b dx f(x) &\approx& \sum_{i=1}^N \phi_i f(x_i) \nonumber \\
\phi_i &=& 
\begin{cases}
\frac{1}{2}(x_2-x_1) & \text{for} \;\;\; i = 1 \\
\frac{1}{2}(x_N-x_{N-1}) & \text{for} \;\;\; i = N \\
\frac{1}{2}(x_{i+1}-x_{i-1}) & \text{for} \;\;\; i = 2,\ldots,N-1 
\end{cases}
\nonumber
\eea
where $N$ is the number of data points and $x_i \in [a,b]$ are the (ordered) discrete observations. 
This is done separately for the calls and the puts. Then, the LHS if Eqs. (\ref{moment1}) - (\ref{moment3}) is fitted by the RHS in the sense of minimum square error. Note that at this step, we take $L_1 =1000$ in our model. In other words our global set of parameters are: $(\alpha_1,\alpha_2,\alpha_3,L_1,L_2,L_3) = (0.1,0.4,0.5,1000,36,6)$.
As mentioned before, the estimate is done sequentially: first we estimate $\lambda_2$ using Eq. (\ref{moment1}), then we proceed to estimate $\lambda_3$ using Eq. (\ref{moment2}). Finally, the kurtosis risk premium is found using Eq. (\ref{moment3}) while enforcing the constraint Eq. (\ref{bound}). 

In figures \ref{fig:vfit1} - \ref{fig:kfit1} we show some examples of varswap, skew and kurtosis fits obtained by calibrating Eqs. (\ref{moment1}) - (\ref{moment3}). We can see that overall, our GARCH model captures quite well the shape of both the varswap and skew term structure. For the kurtosis, the fits are less good. However, we must point out that the moments defined by the LHS of Eq. (\ref{moment3}) are not very stable under the choice of range of deltas. Moreover, we find that the in basically all fits, the constraint given in Eq. (\ref{bound}) is either saturated or very close to being saturated. This means that the kurtosis risk premium is not really independent!
\myfig{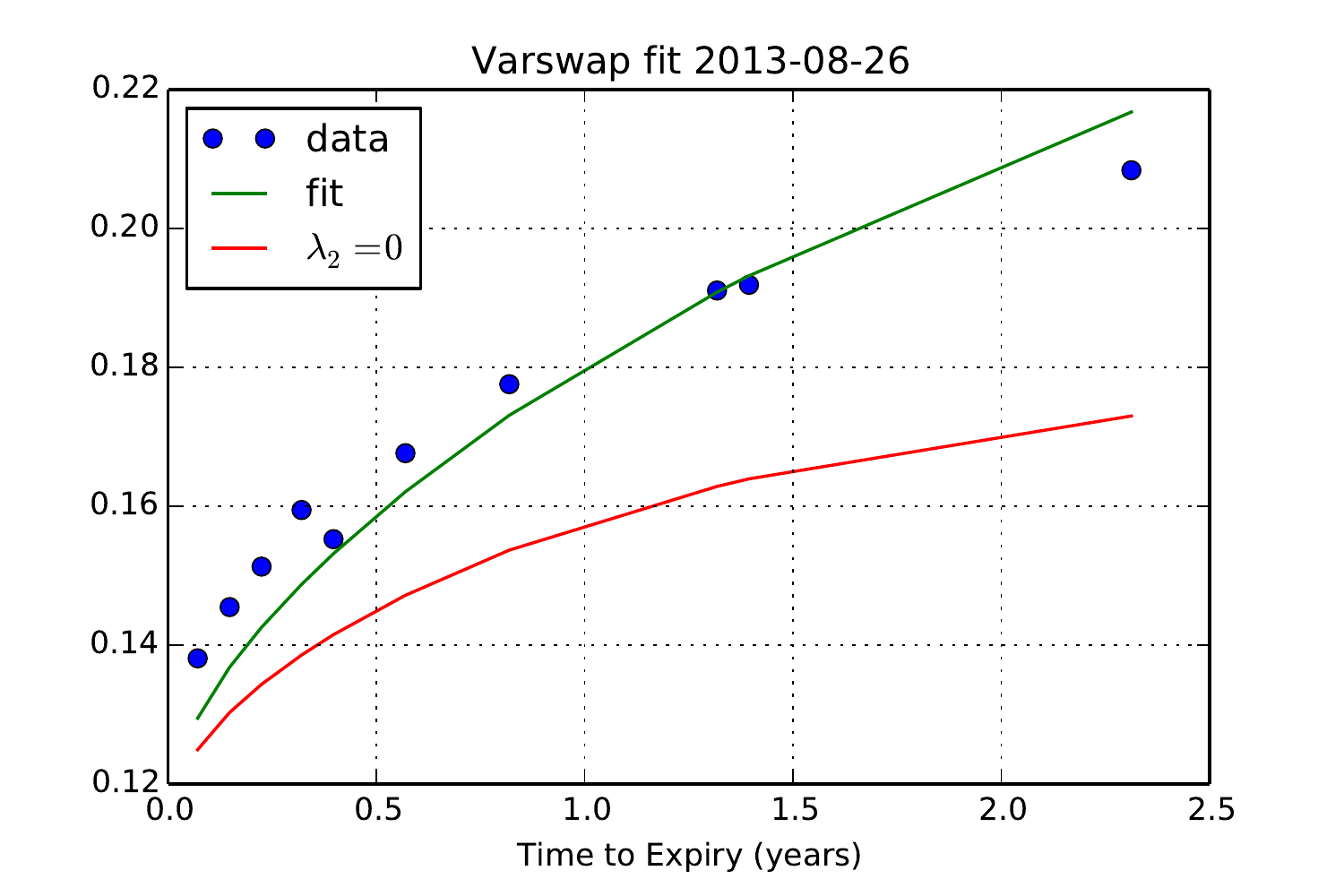}{0.6}{\small GARCH varswap calibration example. The curve with zero risk premium is also shown. Note how the historical estimate of the variance swap level ($\lambda_2=0$) is below the implied one.  }{vfit1}
\myfig{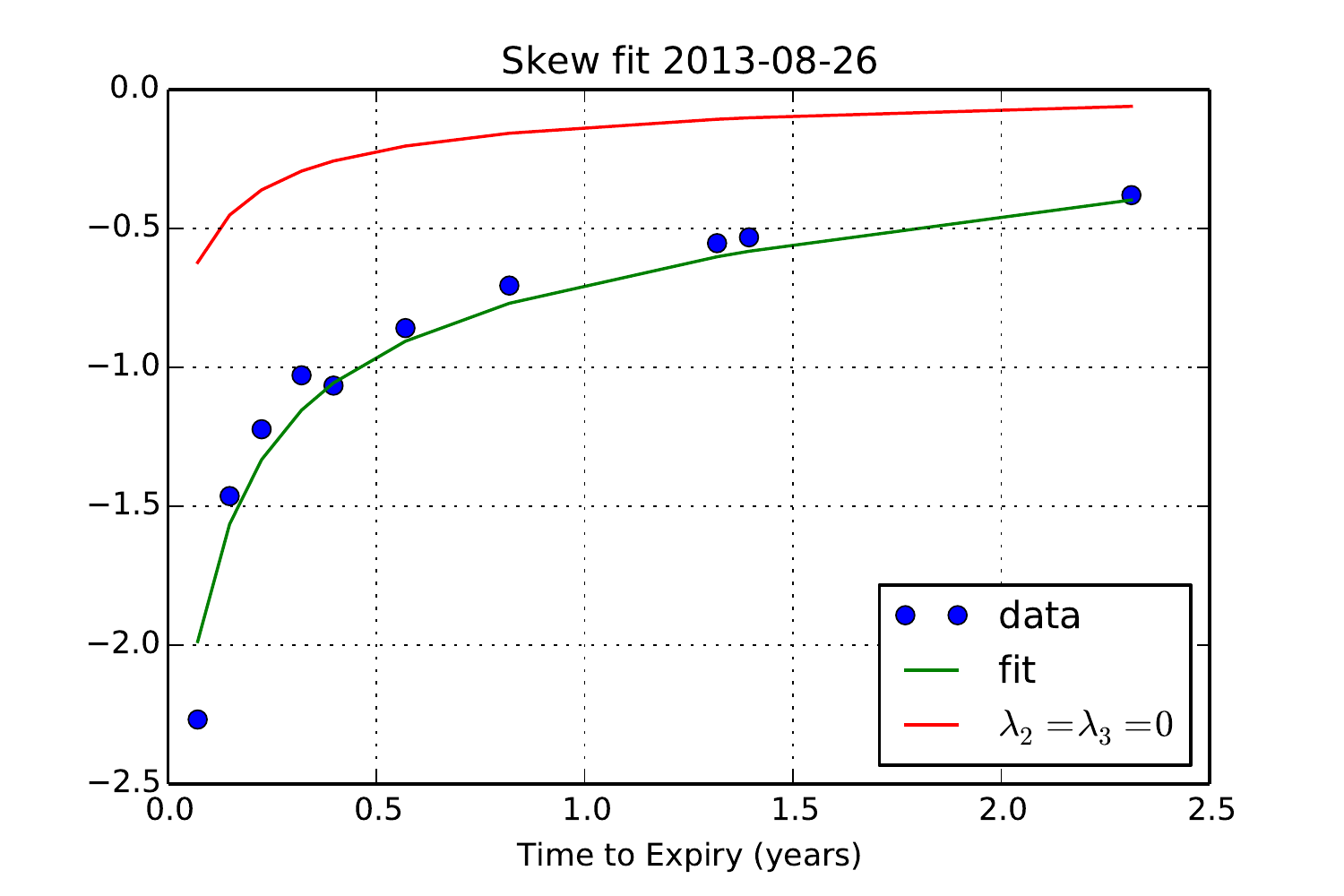}{0.6}{\small GARCH skew calibration example. The curve with zero risk premium is also shown. Without the risk premia we cannot explain the magnitude of the implied skew.}{sfit1}
\myfig{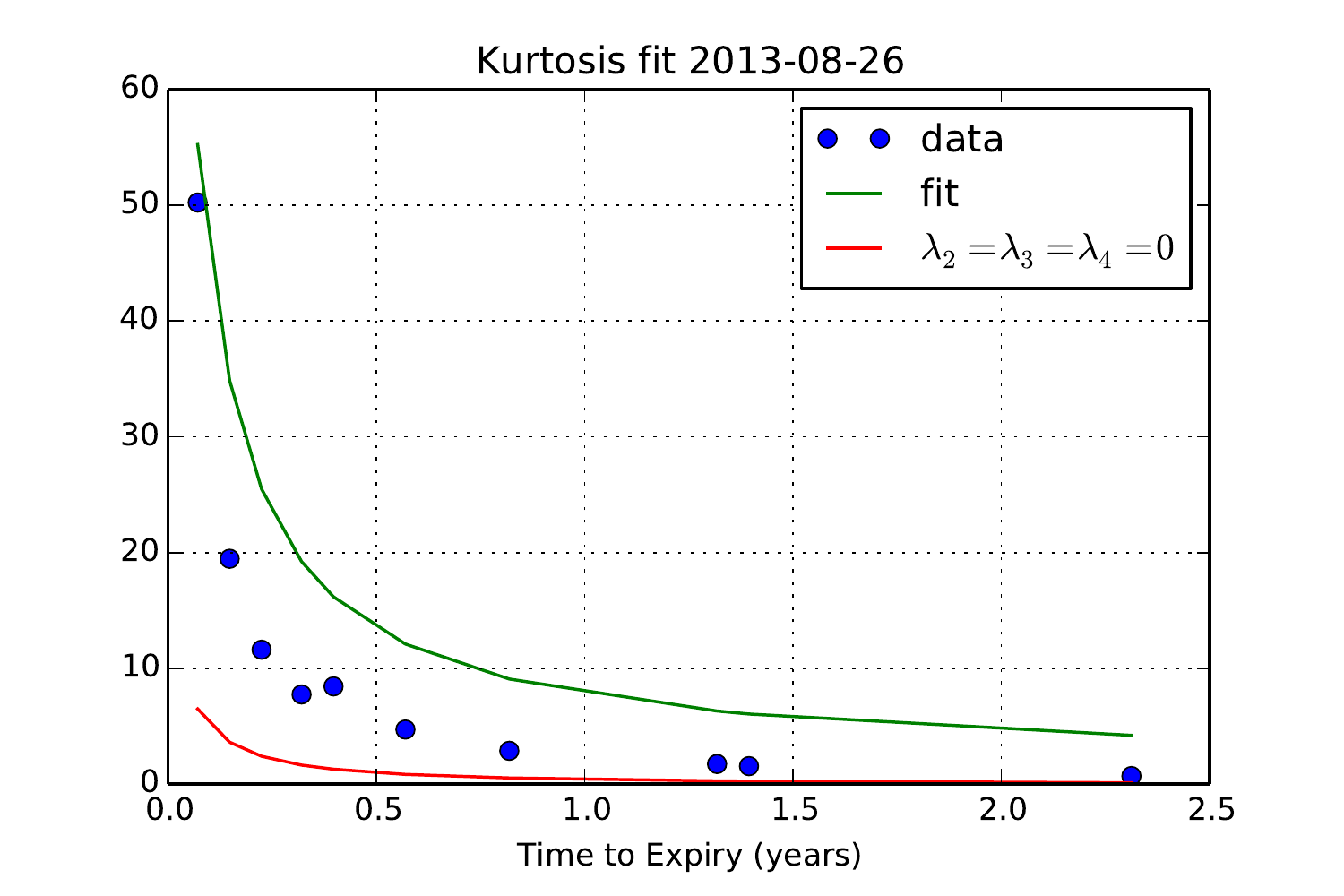}{0.6}{\small GARCH kurtosis calibration example. The curve with zero risk premium is also shown. }{kfit1}

Once we fit the three risk premia, we can generate a full volatility surface by doing Monte Carlo simulations of Eqs. (\ref{mgarchopt}) - (\ref{mgarchrhoij}). To avoid negative prices, we simulate the log returns $d\log S = -\frac{1}{2} (1+ \lambda_2) \nu_t dt + \sqrt{(1+\lambda_2)\nu_t} dW_t$. All equations are then discretized in the standard way, and Brownian motions are simulated using Gaussian innovations. In order to generate the different risk factors, we use the decomposition of Eq. (\ref{dZ}). In figure \ref{fig:mc1} - \ref{fig:mc3} we show some example of volatility smiles. The data is composed of mid-prices of OTM calls and puts with deltas in the following range $|\Delta| \in [0.001,0.999]$. Note that not all fits are very good, however, a bad fit to the varswap term structure does not imply necessarily a bad fit to the overall volatility surface (see figs \ref{fig:vfitbad} - \ref{fig:mc3bad}).
\myfig{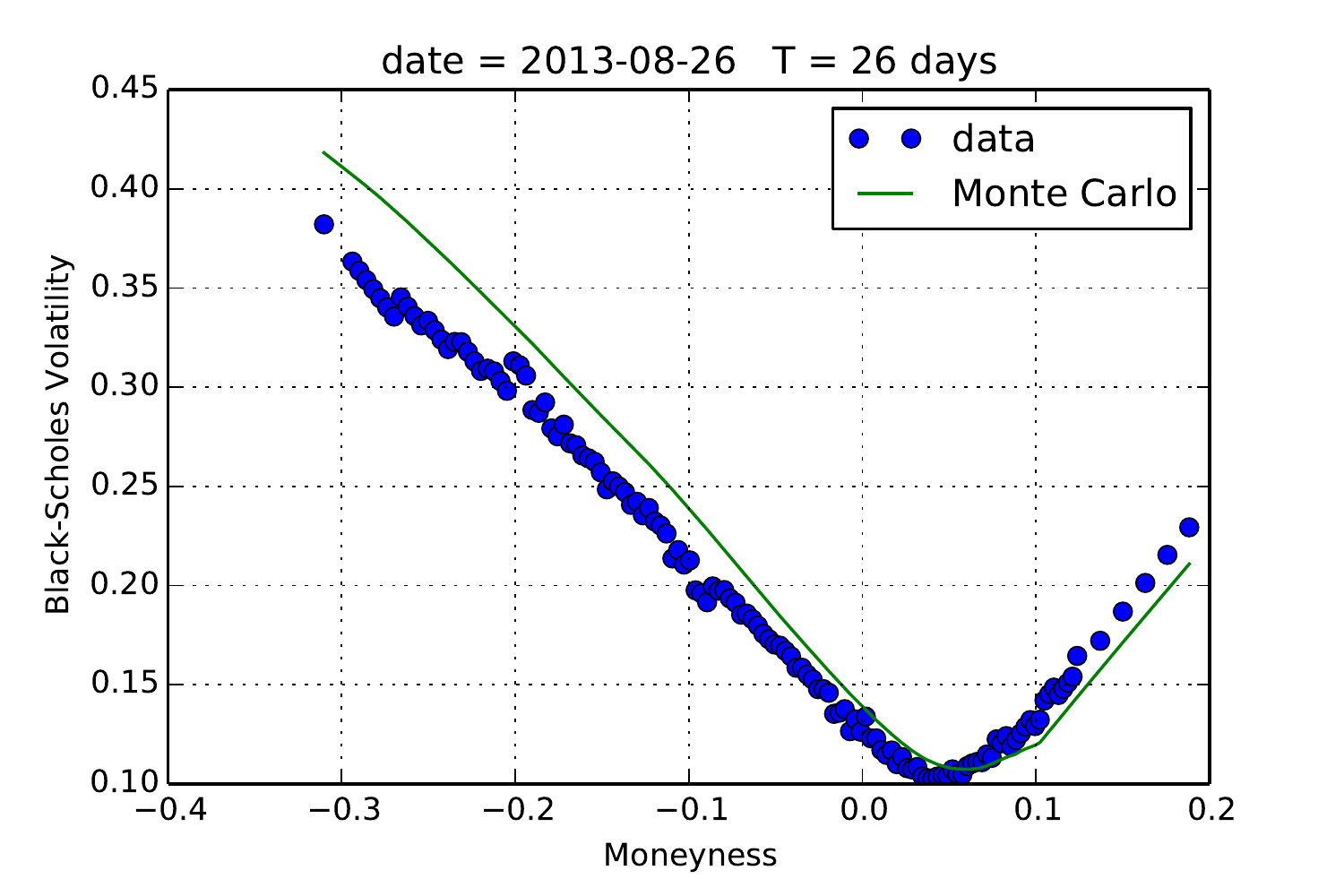}{0.6}{\small Monte Carlo smile fit example.  }{mc1}
\myfig{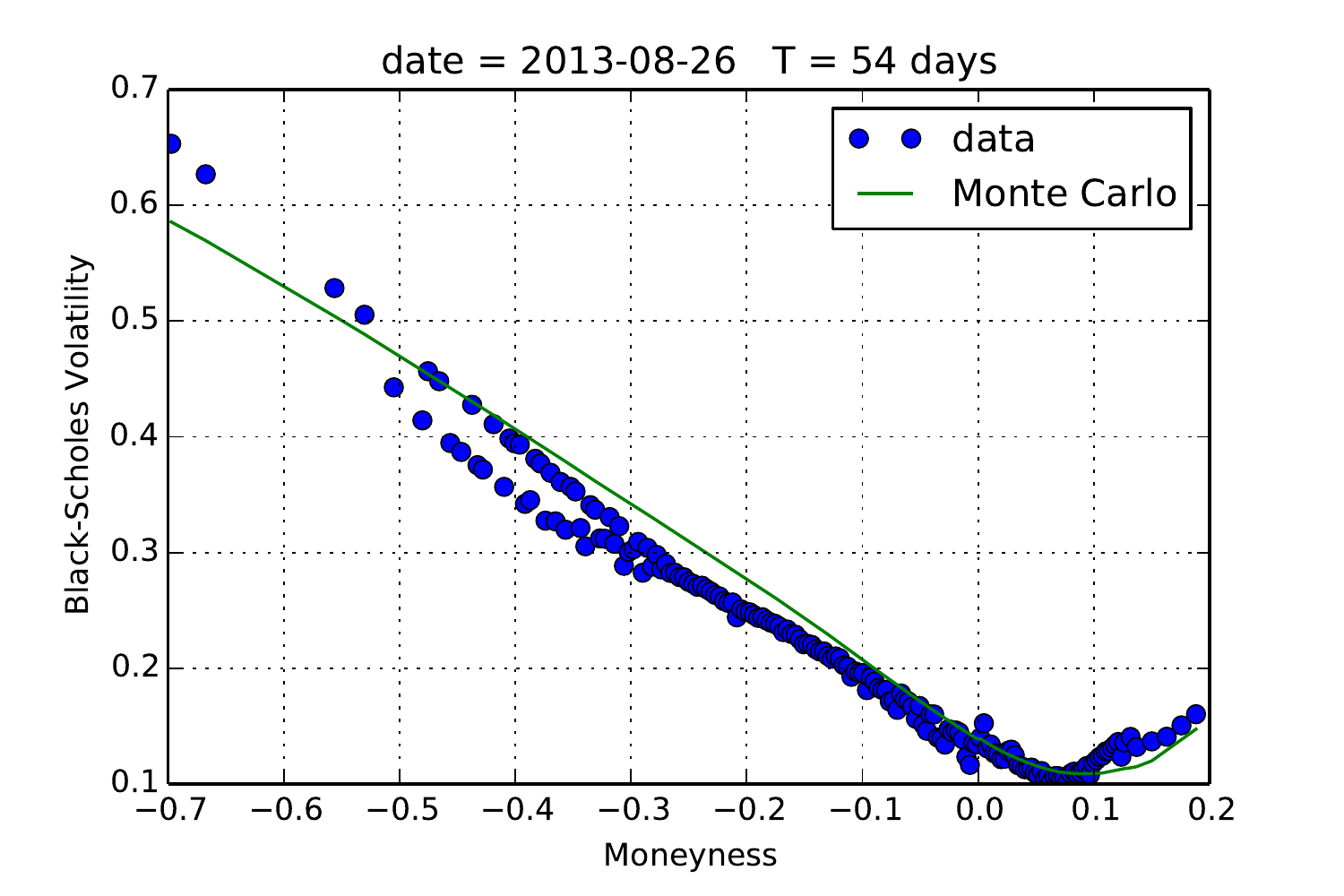}{0.6}{\small Monte Carlo smile fit example.  }{mc2}
\myfig{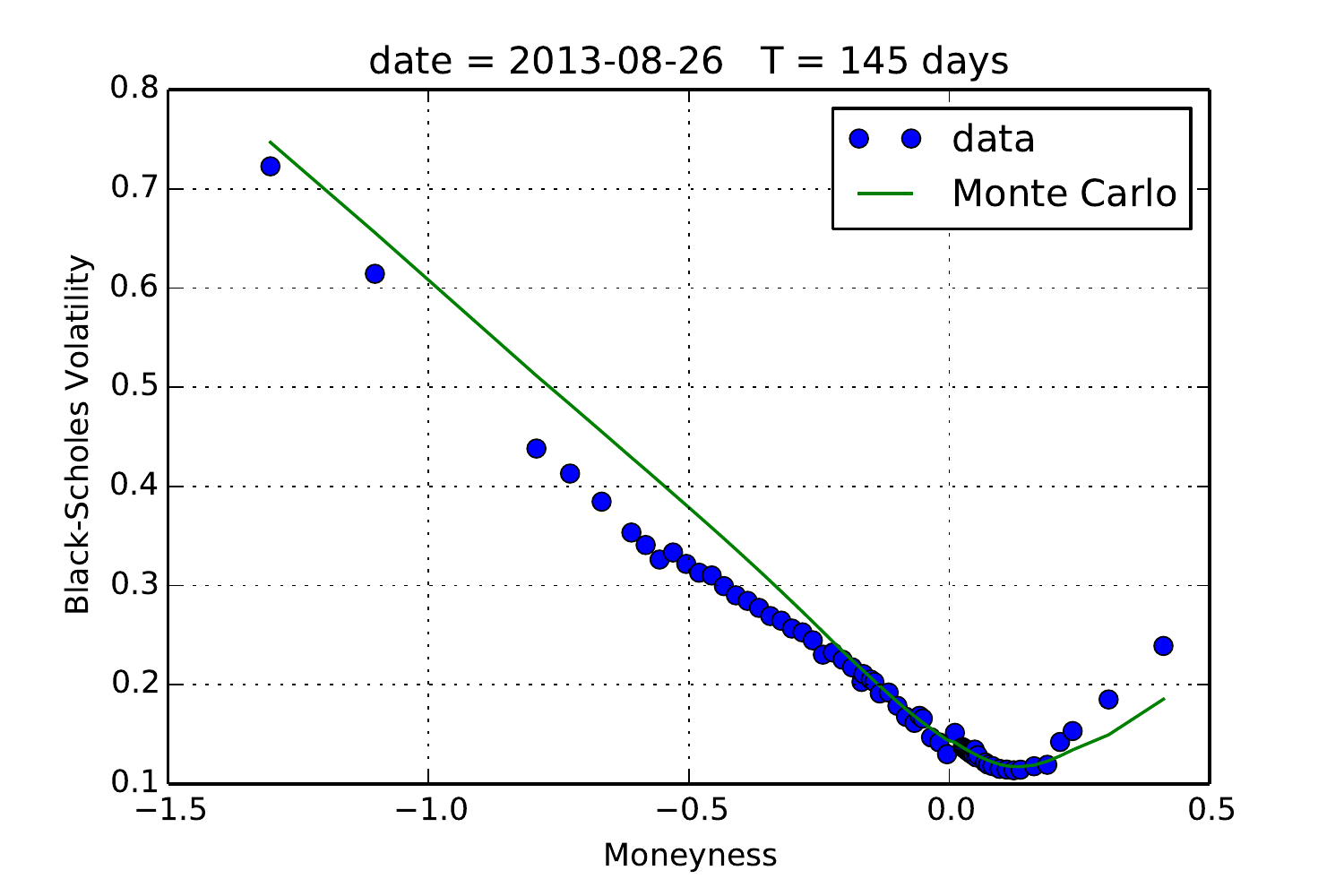}{0.6}{\small Monte Carlo smile fit example.  }{mc3}

\myfig{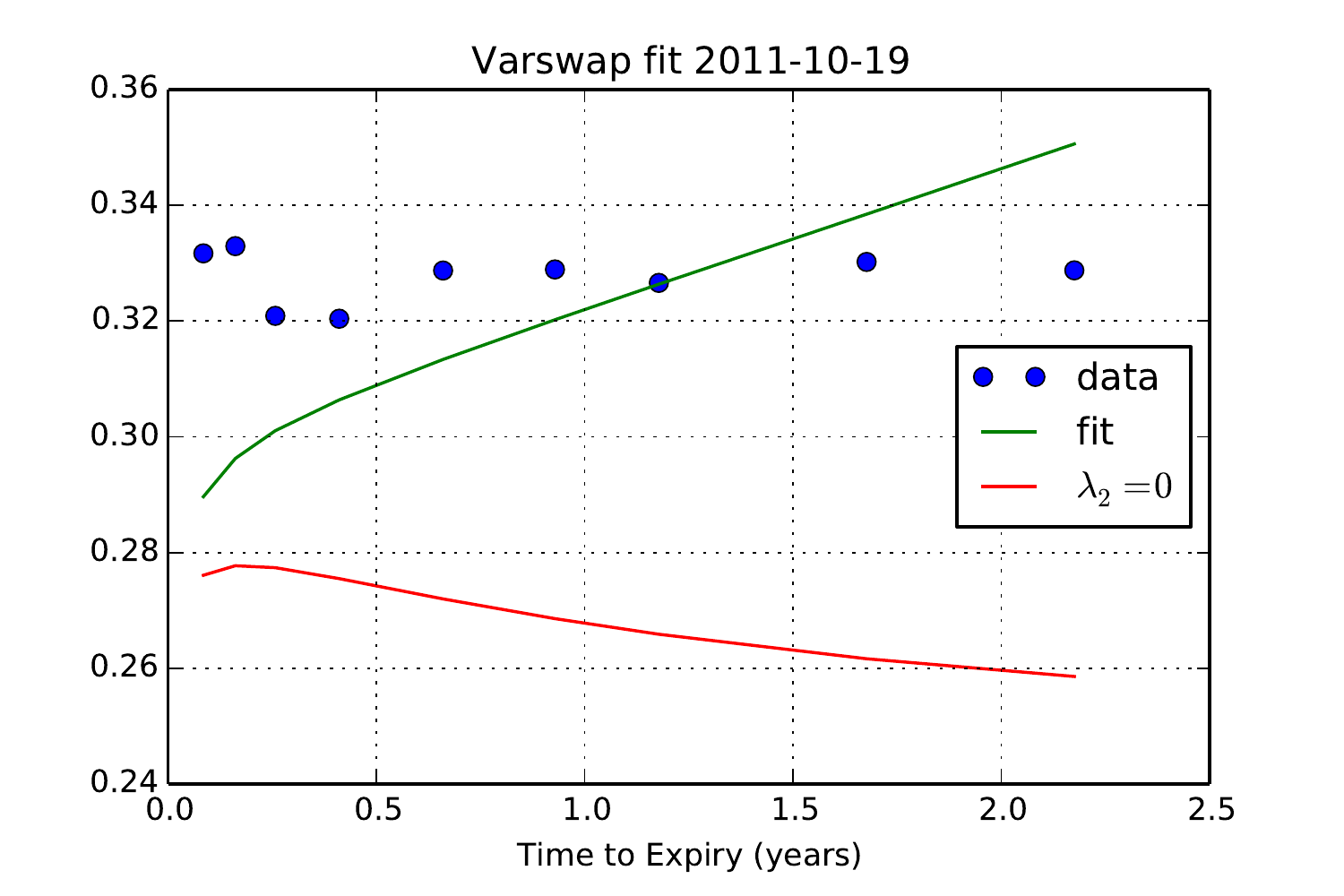}{0.6}{\small A not-so-good varswap fit. }{vfitbad}
\myfig{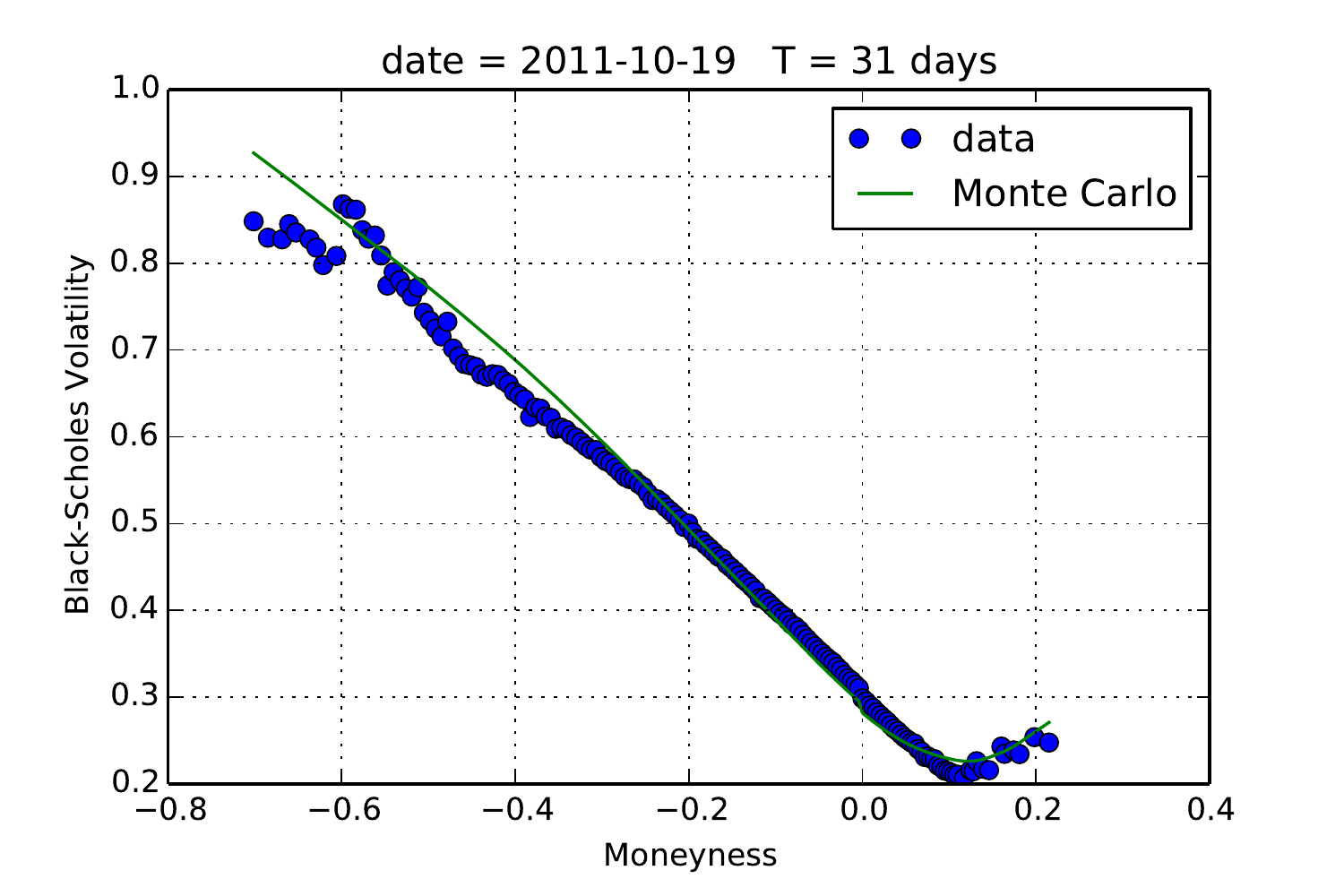}{0.6}{\small Monte Carlo smile fit example when the varswap fit is not good (see figure \ref{fig:vfitbad}).  }{mc1bad}
\myfig{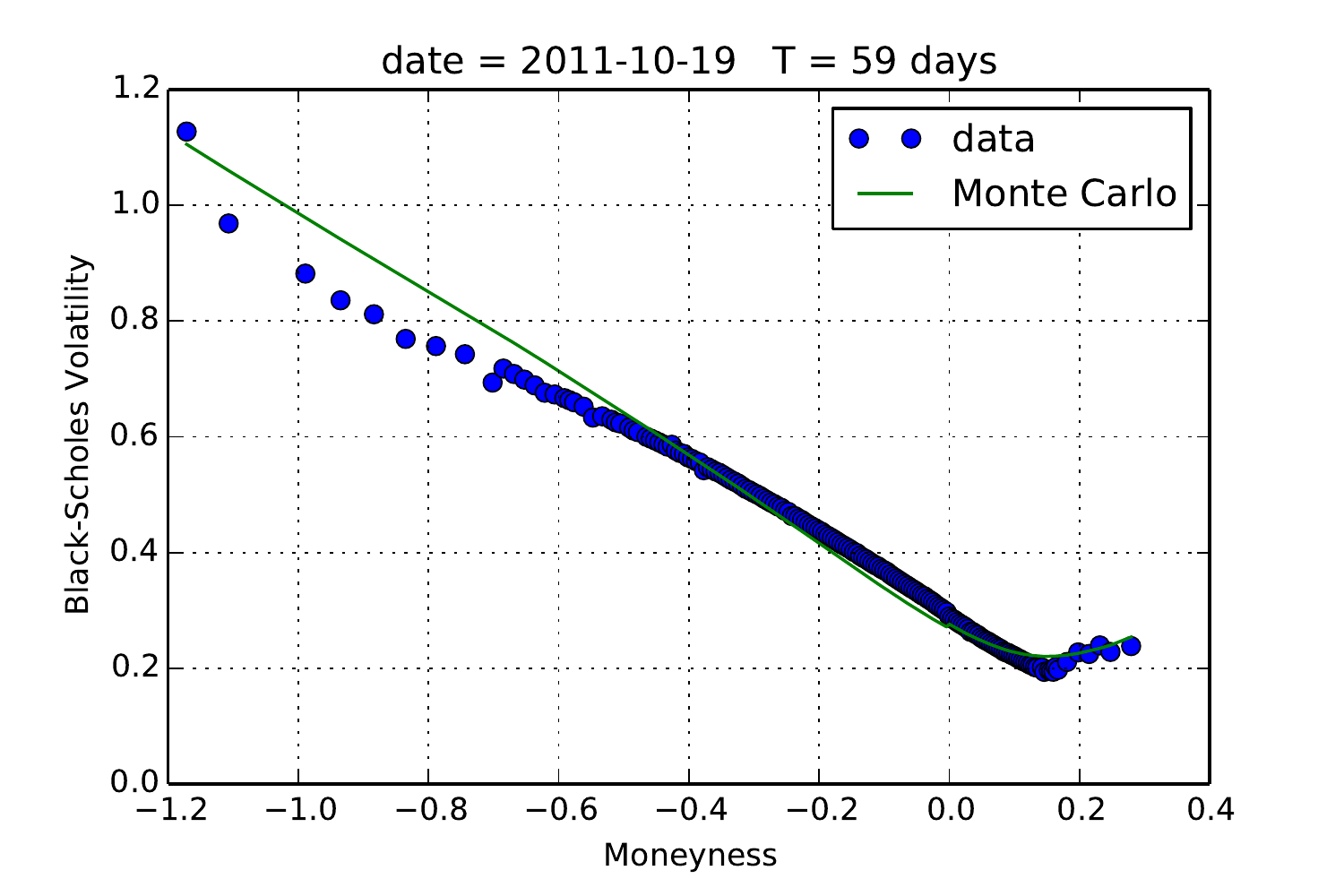}{0.6}{\small Monte Carlo smile fit example when the varswap fit is not good (see figure \ref{fig:vfitbad}).  }{mc2bad}
\myfig{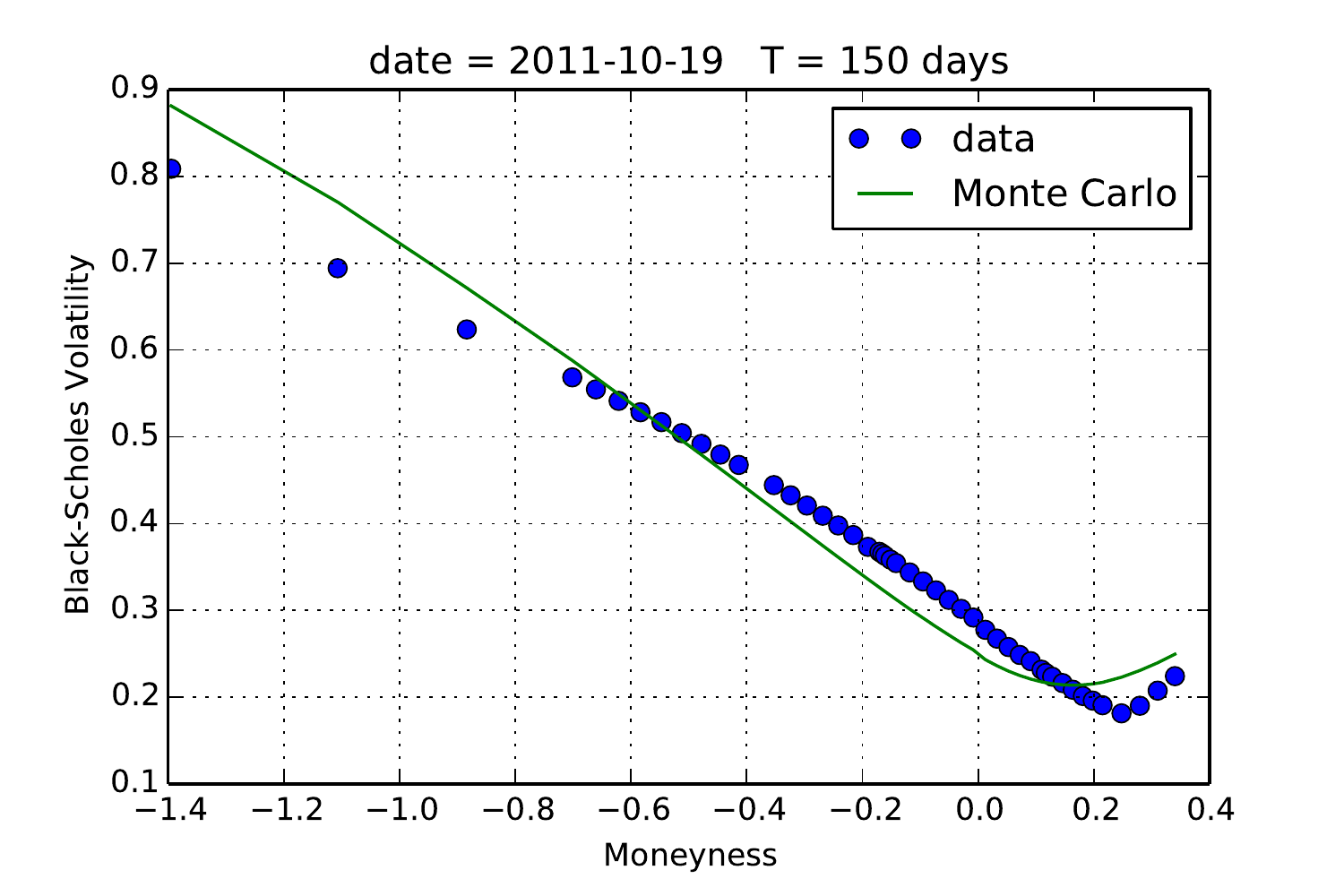}{0.6}{\small Monte Carlo smile fit example when the varswap fit is not good (see figure \ref{fig:vfitbad}).  }{mc3bad}

The time series of the risk premia are shown in figures \ref{fig:lambda2} - \ref{fig:lambda4}. 
One interesting observation is that we find a non-trivial skew risk premium, which is positive and quite stable in time (see fig. \ref{fig:lambda3}). In fact, we can see from figure \ref{fig:sfit1} that when the risk premia are zero, the GARCH model cannot explain the implied skew. Note that even though we set $\mathbb{E}[\epsilon^3] = 0$, we find little evidence for skewness in the distribution of $\epsilon$, so it could never explain the large implied skew. 

The convexity risk premium $\lambda_2$ is not stable in time. It it positive on average, as expected, but it can become negative, specially during a crisis. 
We also note that the kurtosis risk premium $\lambda_4$ saturates the bound given in Eq. (\ref{bound}) most of the time. In fact, the peaks in figure \ref{fig:lambda4} are due to days where the bound was not saturated.  This means, that this is not really an independent parameter, and that our model can be further reduced by imposing equality in Eq. (\ref{bound}). One possible explanation is that our models are too restrictive as all filters are driven by the underlying returns. In fact, for practitioners, we recommend not fitting the kurtosis risk premium and simply saturating the bound of Eq. (\ref{bound}) to find $\lambda_4$.

\myfig{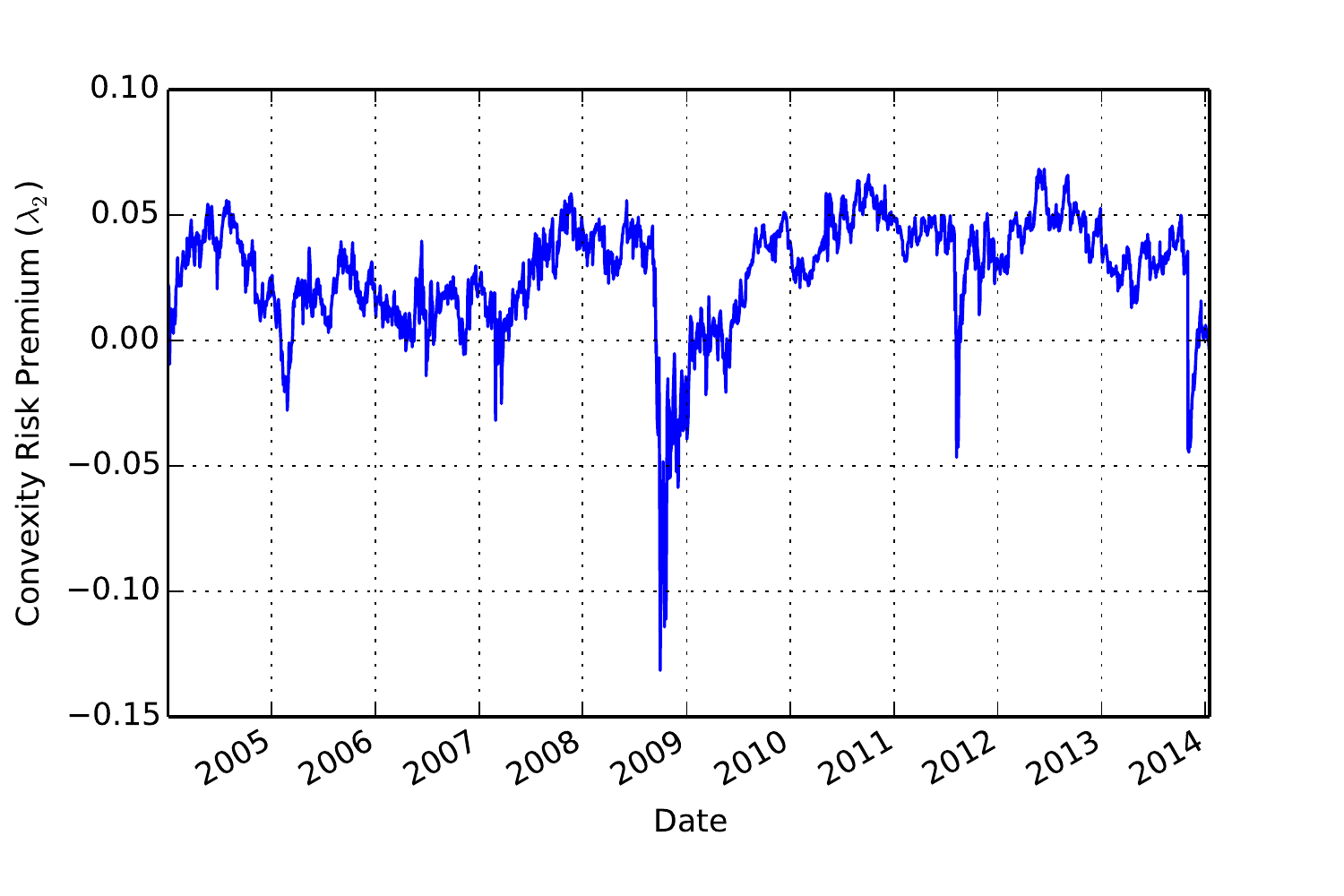}{0.6}{\small Convexity risk premium. }{lambda2}
\myfig{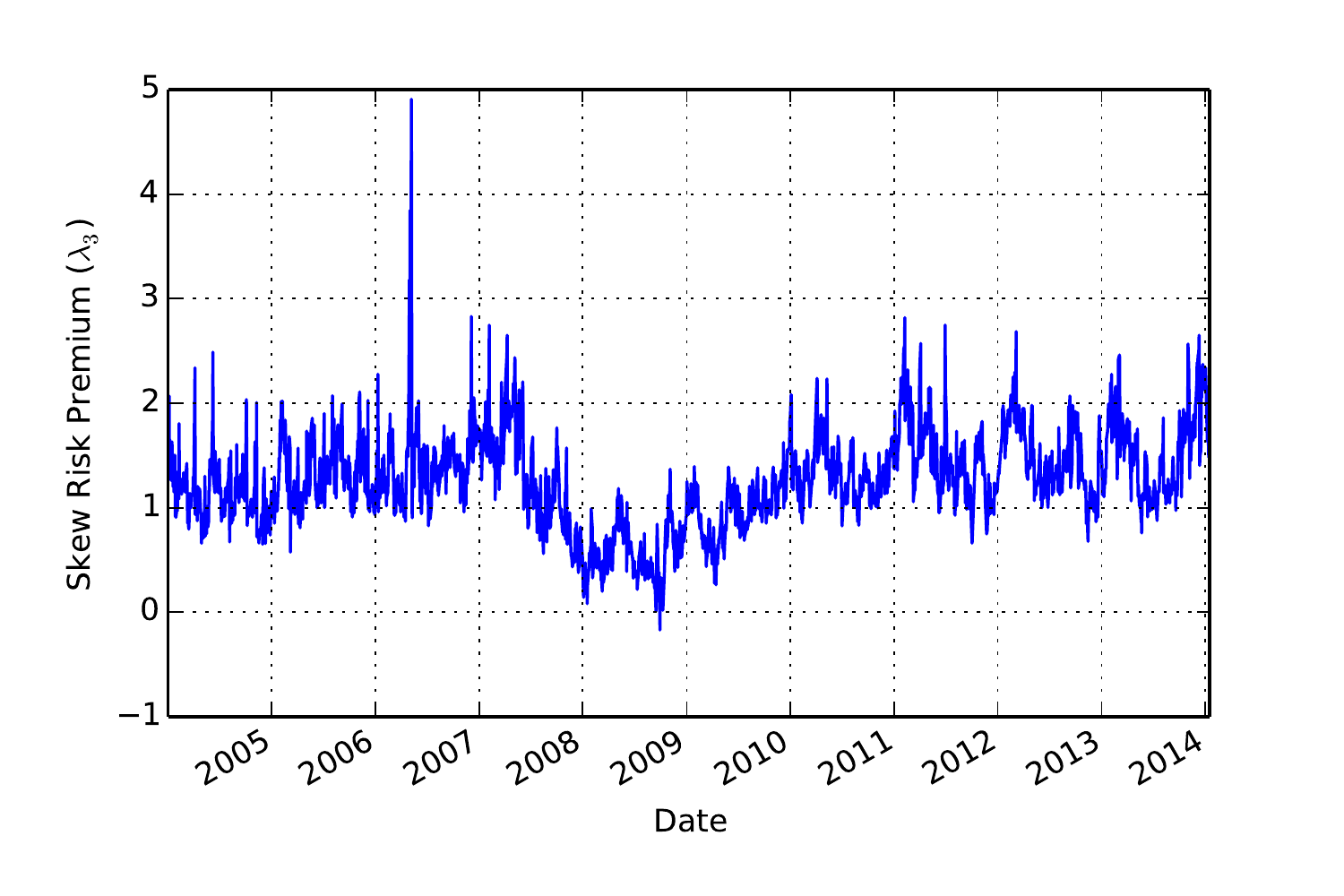}{0.6}{\small Skew risk premium. }{lambda3}
\myfig{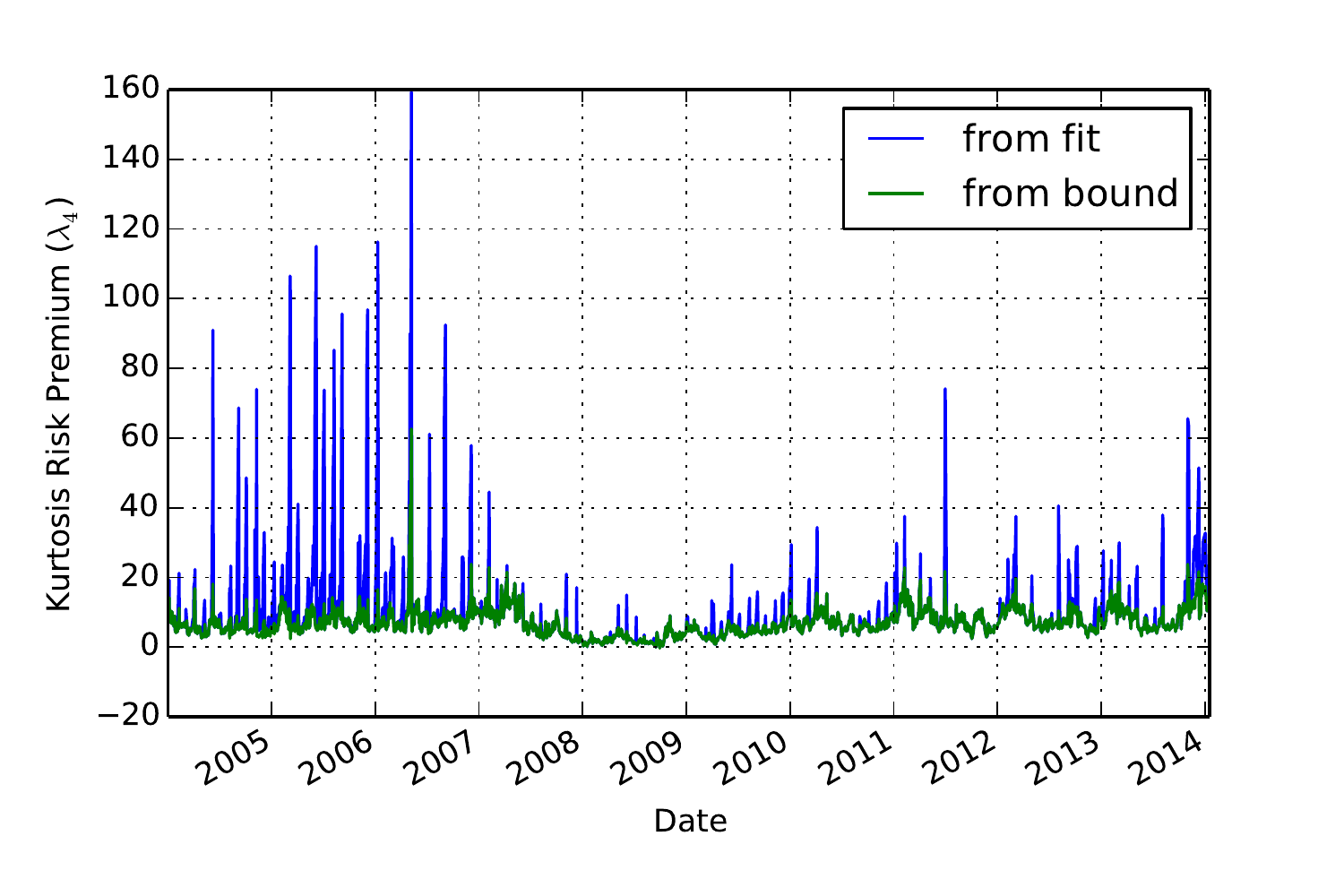}{0.6}{\small Kurtosis risk premium. We show the time series which is obtained by the fit, and the one which comes from saturating the bound of Eq. (\ref{bound}). Notice that most of the
time, both time series are the same. We recommend to simply saturating the bound and avoiding fitting the kurtosis risk premium.}{lambda4}


\section{Conclusion}
\label{sec:conclusion}
In this paper we introduced a new option pricing methodology which uses historical volatility together with risk premium estimates.  In a nutshell, we propose that option prices under the real-world measure are not martingales, but that their drift is governed by tail risk premia. This is because option traders have limited capital, and they face potentially large losses for large market moves due to the non-linear nature of the option contract. 

In particular, we have studied a general class of GARCH models with multiple time scales and asymmetry. However, we believe our procedure can be extended to any kind of volatility estimator, once we can quantify its asymptotic expansion under large movements of the underlying (e.g. tail risks). Within the context of the models studied in this paper, we have found that, if we expand option prices up to second order Greeks, we only need three risk premia: convexity, skew and kurtosis. However, empirically we found that the kurtosis risk premium is not independent and saturates a bound, which allows us to write it in terms of the other two risk premia. Therefore, at the end, our model only has two parameters that must be fitted to option prices: the convexity and skew risk premia. The rest of the parameters are completely determined by historical data using the standard GARCH calibration methodology.
This allows us to generate option smiles which are conditioned on both our volatility forecasts and the market's risk premia. 

We found that the convexity risk premium not only shifts the level of the implied volatility but also changes its term structure. On the other hand, the skew risk premium makes the ATM skew steeper (more negative) than the historical estimate, and the kurtosis risk premium makes the vol-of-vol higher than the historical one. 

We developed a calibration methodology based on the vol-of-vol expansion of Bergomi and Guyon \cite{bergomi}, where we fit a series of implied moments which can be replicated using OTM options. We then derived approximate formulas for these moments up to second order in vol-of-vol. Once the risk premia are found, we can generate the full volatility surface by using Monte Carlo simulations. We showed that the smiles obtained this way are reasonably close to what is observed in the SPX option market.

There are various extensions to our work that deserve more attention.
In particular, we have assumed that we can approximate option prices to second order Greeks even though we work in discrete time. It would be interesting to relax this assumption. Perhaps this can be done in the context of the Hedged Monte Carlo method of \cite{hmc}. Another extension of our model is to make the convexity risk premium time dependent. As we saw in section \ref{sec:calibration}, the skew and kurtosis risk premia are quite stable in time. However, this is not the case for the convexity risk premium, which can even become negative during a crisis. Finally, it would be interesting to explore the relation between our approach and the so-called pricing kernel \cite{jacobs2,jacobs3}.

\section*{Acknowledgments}
I would like to thank Baruch University where part of this research was done. I also thank Arthur Berd, Jim Gatheral, Peter Carr, Adela Baho, Tai Ho Wang, Anja Richter, Andrew Lesniewski and Filippo Passerini for many stimulating discussions and comments on the manuscript.

\end{document}